\documentclass[aps,prd,10pt,twocolumn,superscriptaddress,showpacs,longbibliography,preprintnumbers]{revtex4-1}

\usepackage{aas_macros}
\usepackage{hyperref}
\usepackage{graphicx}
\usepackage{amsfonts,amsmath,amssymb,bm}
\usepackage{array}
\usepackage{xcolor}
\usepackage{comment}
\usepackage{soul}

\newcolumntype{P}[1]{>{\centering\arraybackslash}p{#1}}

\graphicspath{{figures/}}

\begin{document}


\title{Exciting Prospects for Detecting Late-Time Neutrinos from Core-Collapse Supernovae}

\author{Shirley Weishi Li}
\email{shirleyl@slac.stanford.edu}
\thanks{\href{http://orcid.org/0000-0002-2157-8982}{0000-0002-2157-8982}}
\affiliation{Center for Cosmology and AstroParticle Physics (CCAPP), Ohio State University, Columbus, OH 43210}
\affiliation{Department of Physics, Ohio State University, Columbus, OH 43210}
\affiliation{SLAC National Accelerator Laboratory, Menlo Park, CA, 94025}

\author{Luke F. Roberts}
\email{robertsl@nscl.msu.edu}
\thanks{\href{http://orcid.org/0000-0001-7364-7946}{0000-0001-7364-7946}}
\affiliation{National Superconducting Cyclotron Laboratory and Department of Physics and Astronomy, Michigan State University, East Lansing, MI 48824}

\author{John F. Beacom}
\email{beacom.7@osu.edu}
\thanks{\href{http://orcid.org/0000-0002-0005-2631}{0000-0002-0005-2631}}
\affiliation{Center for Cosmology and AstroParticle Physics (CCAPP), Ohio State University, Columbus, OH 43210}
\affiliation{Department of Physics, Ohio State University, Columbus, OH 43210}
\affiliation{Department of Astronomy, Ohio State University, Columbus, OH 43210}

\date{December 7, 2020}

\begin{abstract}
The importance of detecting neutrinos from a Milky Way core-collapse supernova is well known.  An under-studied phase is proto-neutron star cooling.  For SN 1987A, this seemingly began at about 2 s, and is thus probed by only 6 of the 19 events (and only the $\bar{\nu}_e$ flavor) in the Kamiokande-II and IMB detectors.  With the higher statistics expected for present and near-future detectors, it should be possible to measure detailed neutrino signals out to very late times.  We present the first comprehensive study of neutrino detection during the proto-neutron star cooling phase, considering a variety of outcomes, using all flavors, and employing detailed detector physics.  For our nominal model, the event yields (at 10~kpc) after 10~s---the approximate duration of the SN 1987A signal---far exceed the entire SN 1987A yield, with $\simeq$250 $\bar{\nu}_e$ events (to 50~s) in Super-Kamiokande, $\simeq$110 $\nu_e$ events (to 40~s) in DUNE, and $\simeq$10 $\nu_\mu, \nu_\tau, \bar{\nu}_\mu, \bar{\nu}_\tau$ events (to 20~s) in JUNO.  These data would allow unprecedented probes of the proto-neutron star, including the onset of neutrino transparency and hence its transition to a neutron star.  If a black hole forms, even at very late times, this can be clearly identified.  But will the detectors fulfill their potential for this perhaps once-ever opportunity for an all-flavor, high-statistics detection of a core collapse? Maybe.  Further work is urgently needed, especially for DUNE to thoroughly investigate and improve its MeV capabilities.
\end{abstract}

\preprint{SLAC-PUB-17548}

\maketitle


\section{Introduction}
\label{sec:intro}

For massive stars, core collapse is inevitable, though the details are uncertain~\cite{Arnett:1990au, Janka:12, Scholberg:2012id, Kotake:2012iv, Burrows:13, Mirizzi:2015eza, Smartt:2015sfa, Janka:16, Mueller:16}.  For a successful supernova, core collapse creates a hot proto-neutron star (PNS) and launches a shock wave that propagates through the stellar envelope, leading to the characteristic optical display.  As the PNS cools, it typically becomes a neutron star (NS), though black-hole (BH) formation is possible through a phase transition in the PNS or fallback onto it.  For a failed supernova the shock wave is too weak, and it reverses, causing the whole star to form a BH with no to little optical display.  Probing core collapse is important to understanding massive-star fates, hot nuclear matter, NSs and BHs, and the products of successful supernovae, which include chemical elements, cosmic rays, electromagnetic transients and energetic material that drives galactic feedback.  (Properly, ``core collapse" means either a successful or failed explosion and ``supernova" means a successful one, but the latter is sometimes used to mean both.)

Neutrinos are critical to understanding core collapse.  Only neutrinos can radiate away the huge changes in gravitational binding energy and lepton number. (Detectable gravitational-wave signals can also emerge from the PNS, but are produced only if there is sufficient deviation from spherical symmetry and carry little energy.)  The change in gravitational binding energy of the $\simeq$1.4$M_\odot$ iron core as it becomes a $\simeq$12-km NS is $- \Delta E_{\rm B} \simeq (3/5) G_{\rm N} M_{\rm NS}^2 / R_{\rm NS} \simeq 3 \times 10^{53}$ erg, which is $\simeq$10\% of its rest-mass energy.  The change in lepton number due to the conversion of the iron core into a NS is $M_{\rm NS}/2m_p \simeq 8 \times 10^{56}$, and it is carried by a number excess of $\nu_e$ relative to $\bar{\nu}_e$.  Although neutrinos are temporarily trapped in the PNS and must diffuse out, they readily pass through the surrounding stellar envelope, which electromagnetic radiation cannot do, and provide a probe of the supernova central engine.  Soon after explosion, their average energies are $\simeq$10~MeV for $\nu_e$, $\simeq$13~MeV for $\bar{\nu}_e$, and $\simeq$14~MeV for $\nu_\mu, \nu_\tau, \bar{\nu}_\mu, \bar{\nu}_\tau$ (hereafter collectively called $\nu_x$), inversely reflecting the strengths of their interactions with the neutron-rich matter of the PNS. At later times, the average neutrino energies decrease and converge to each other.

We must detect neutrinos from a Milky Way core collapse and we must detect them well.  For SN 1987A (in the Large Magellanic Cloud), we have only 19 events in the Kamiokande-II~\cite{Hirata:1987hu, Hirata:1988ad} and Irvine-Michigan-Brookhaven (IMB)~\cite{Bionta:1987qt, Bratton:1988ww} detectors---and only the $\bar{\nu}_e$ flavor---over $\simeq$10~s.  This was enough to probe the basics of core collapse and to test some neutrino properties, but key questions remain unanswered.  For the next Milky Way core collapse---likely 5 times closer than SN 1987A---we will have much better detectors, with Super-Kamiokande (Super-K)~\cite{Abe:2016nxk} leading on detecting $\bar{\nu}_e$, the Deep Underground Neutrino Experiment (DUNE)~\cite{Abi:2020evt} on $\nu_e$, and the Jiangmen Underground Neutrino Observatory (JUNO)~\cite{An:2015jdp} on $\nu_x$.  (For $\bar{\nu}_e$, JUNO and eventually Hyper-Kamiokande (Hyper-K)~\cite{Abe:2018uyc} will also be excellent.)  We focus on the detectors with the largest numbers of identifiable events at late times; for others, see Ref.~\cite{Scholberg:2012id}.

Because core collapses are so rare (a few per century in the Milky Way and its satellites~\cite{Reed:2005en, Diehl:2006cf, Keane:2008jj, 2011MNRAS.412.1473L, Adams:2013ana,Rozwadowska:2021lll}, the maximum range for detectable neutrino bursts), it is essential that we make complete measurements.  {\it We may have only one chance to detect a core collapse with high precision in all neutrino flavors.}  The present and planned huge neutrino detectors are designed primarily to measure mixing using terrestrial sources, and are not fully optimized to detect core collapses.  Theory work is needed now to define expectations, assess readiness, and suggest improvements.  Further, once the neutrino-mixing missions of these detectors are achieved, it is not clear if all of them (or any successors) will run long enough to detect a Milky Way core collapse.  Without the full flavor coverage of this complement of detectors, our ability to probe core-collapse physics would be significantly degraded.

It is important to detect neutrinos to the latest possible times.  This will probe PNS physics in detail and accurately measure the total radiated energy and lepton number.  In nominal models, the physics beyond a few seconds is dominated by PNS cooling, with increasingly similar emission in all flavors.  By ``late-time" emission, we mean the late PNS-cooling phase, which may begin well before 10 s, as discussed in Sec.~\ref{sec:intro_CCSN}.  After a few tens of seconds, the PNS becomes neutrino-transparent, leading to a rapid drop in the fluxes, marking the formation of a NS.  But there are other possible outcomes, including BH formation, which would sharply truncate the flux, and which could occur early or late~\cite{Burrows:1988ba, Baumgarte:1996iu, Beacom:2000qy, Nakazato:2007yd, Sumiyoshi:2007pp, Fischer:2008rh, OConnor:2010moj, Kuroda:2018gqq, Schneider:2020kxr}.  For SN 1987A, the low statistics beyond 2 s---only 6 of the 19 events, and all $\bar{\nu}_e$---make it hard to measure the physics of NS formation or to test for more exotic outcomes.  The fate of the SN 1987A's collapsed core is unknown~\cite{Graves:05, Manchester:2007be, Alp:2018oek, Esposito:2018nib, Zhang:2018dez, Cigan:2019shp, Page:2020gsx}, showing the importance of better neutrino measurements.

\begin{figure}[t]
\centering
\includegraphics[width=\columnwidth]{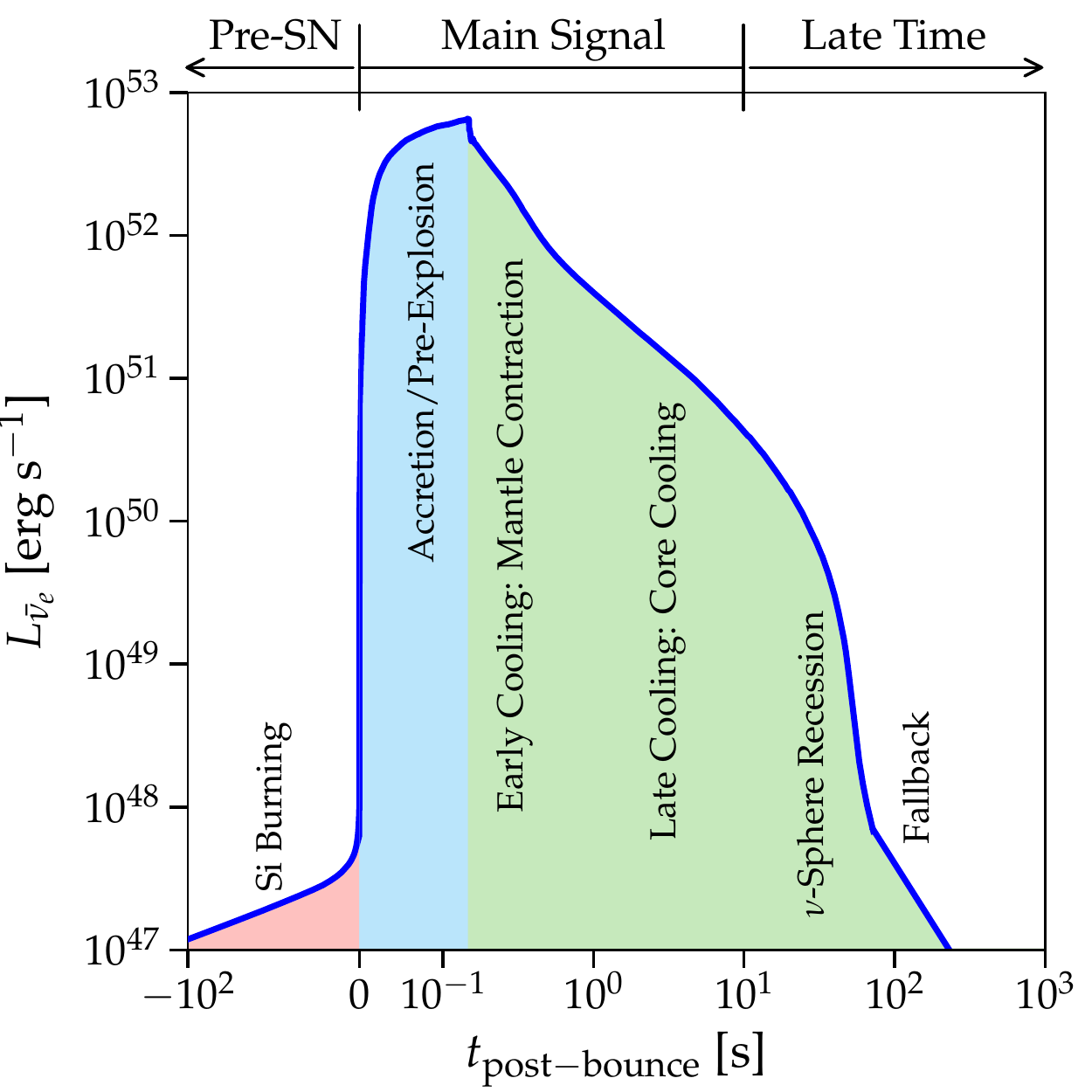}
\caption{Schematic illustration of the $\bar{\nu}_e$ emission profile from a successful core-collapse supernova.  The time axis is linear before 0~s, linear from 0 to $10^{-1}$~s with a different scale, and logarithmic after $10^{-1}$~s.  The different physical phases---pre-SN (red), accretion/pre-explosion (blue), and cooling (green)---are shaded, with key periods noted.  The labels on the top axis show common---but not physically motivated---descriptions.}
\label{fig:illustration}
\end{figure}

{\it In this paper, we present the first comprehensive study of PNS-cooling neutrino signal detection from core collapse, highlighting late times.}  We improve upon earlier work~\cite{Keil:95, Pons:99, Pons:01a, Roberts:12, Nakazato:2012qf, Mirizzi:2015eza, Camelio:17, Suwa:2019svl, Nakazato:2020ogl} by providing a complete conceptual framework and by calculating results for all flavors, emphasizing spectra, and using detailed detection physics. Many considerations make this timely: Super-K is adding dissolved gadolinium, the design of DUNE is being finalized, and JUNO's construction is nearly done.  Our goals are to frame and highlight the physics opportunities of PNS-cooling neutrino detection, to motivate improvements to experiments, and to encourage further simulation and phenomenological work.  Overall, our results---which include new quantitative assessments of flavor coverage, time profiles, spectra, and uncertainties---show that the late-time frontier is very promising.

\renewcommand{\arraystretch}{1.5}
\begin{table}
\centering
\begin{tabular}{||c||P{6.1cm}||}
\hline
\textbf{Phase} & \textbf{Physics Opportunities} \\
\hline\hline
Pre-SN & early warning, progenitor physics \\ \hline
Neutronization & flavor mixing, SN distance, new physics \\ \hline
Accretion & flavor mixing, SN direction, multi-D effects \\ \hline
Early cooling & equation of state, energy loss rates,
\par PNS radius, diffusion time, new physics \\ \hline
Late cooling & NS vs. BH formation, transparency time,
\par integrated losses, new physics \\ \hline
\end{tabular}
\caption{Key physics opportunities from detecting supernova neutrinos in different phases.}
\label{tab:physics}
\end{table}

In the following, we begin by reviewing the physics behind neutrino emission and detection (Sec.~\ref{sec:intro_CCSN}) as well as the details of the PNS simulation we use (Sec.~\ref{sec:theory}).  We then calculate detection signals for all flavors in the PNS case (Sec.~\ref{sec:resultsNS}), interpret the physics prospects for the PNS and BH cases (Sec.~\ref{sec:physicsresults}), and conclude (Sec.~\ref{sec:conclusions}).


\section{Overview of Core Collapse and Neutrino Emission}
\label{sec:intro_CCSN}

In this section, we provide a conceptual framework for the results and discussions that follow.  We cover the case of a successful core-collapse supernova, focusing on its underlying physics and consequent neutrino emission---from the explosion phase to the PNS cooling phase and then to other possible late-time emission mechanisms---followed by discussions of the effects of neutrino mixing and the needs for neutrino detection.  Figure~\ref{fig:illustration} is a schematic overview of the $\bar \nu_e$ emission profile, with Table~\ref{tab:physics} highlighting key physics opportunities, as described in detail below.  Last, we summarize how a BH may form.


\subsection{Pre-explosion: PNS formation}
\label{subsec:explosion}

Core-collapse supernovae are among the most spectacular, complex events in the Universe.  All four fundamental forces play important roles and many different time and length scales are coupled.  Although many details of core collapse are not fully understood, there is a generally agreed-upon picture that most core-collapse supernovae are powered by  the delayed neutrino heating mechanism~\cite{Bethe:85}, in which the stalled shock is revived by neutrino heating. Here we briefly discuss the various phases of a core-collapse supernova, focusing on the aspects most relevant to neutrino emission~\cite{Arnett:1990au, Janka:12, Scholberg:2012id, Kotake:2012iv, Burrows:13, Mirizzi:2015eza, Smartt:2015sfa, Janka:16, Mueller:16}.

Massive stars ($8 M_\odot \lesssim M \lesssim 100 M_\odot$) are powered by fusion reactions in their cores.  These reactions proceed from the fusion of light elements, e.g., hydrogen and helium, to heavier elements, e.g., carbon and oxygen, culminating with the fusion of silicon isotopes into iron-peak elements, dominantly nickel, beyond which net energy is no longer generated through nuclear reactions~\cite{Woosley:02}.  Throughout this evolution, the star is producing neutrinos through nuclear fusions and beta decays, as well as thermal pair emission, though the fluxes are far too small to measure.  The only potentially detectable flux occurs during silicon burning.  For the closest pre-supernova stars, within 600~pc, the thermal pair flux during silicon burning may be detectable for a few hours~\cite{Odrzywolek:2003vn, Odrzywolek:2010zz, Asakura:2015bga, Patton:2015sqt, Kato:2017ehj, Simpson:2019xwo, Raj:2019wpy}.  We use the flux from Ref.~\cite{Odrzywolek:2010zz} in Fig.~\ref{fig:illustration}.

As the iron core grows, its main support against gravity comes from electron degeneracy pressure.  When its mass exceeds the effective Chandrasekhar mass---which can differ from 1.4$M_\odot$ due to the effects of thermal pressure, the surrounding envelope, and neutronization---then the core starts to collapse.  The inner core collapses homologously while the outer core collapses supersonically. Electrons capture on nuclei and produce $\nu_e$, which gives rise to a rapidly increasing $\nu_e$ luminosity until the core reaches densities where neutrinos become trapped ($\rho \simeq 10^{11} \, \textrm{g/} \textrm{cm}^3$).  When the central density of the core reaches nuclear densities, the stiffening of the equation of state causes the core to bounce.  This defines $t_{\rm{post-bounce}} = 0$ in our figures.  When the bounce disturbance encounters the supersonically infalling material in the outer core, it turns into a shock wave.

The shock wave propagates outward, sweeping through the outer core, heating the material as it passes through. Within a few milliseconds, the shock moves to a position in the star where neutrinos are not trapped and ``breaks out'' of the neutrinosphere. Heating due to the shock drastically increases the rate of neutronization via $e^- + p \rightarrow \nu_e + n$ and gives rise to the ``neutronization burst'' in the $\nu_e$ luminosity around 10~ms after bounce. (It is not included in Fig.~\ref{fig:illustration} because only the $\bar{\nu}_e$ flux is shown.) Importantly, despite its name, this burst carries away only $\simeq$1--3\% of the total energy release and $\simeq$40\% of the total lepton-number release, with the remainders emitted over tens of seconds.  In addition, even the full emission of lepton number accounts for only $\simeq$5\% of the total number of neutrinos emitted.  After breakout, the shock wave loses energy by neutrino emission and by dissociating nuclei. These losses eventually cause the shock to stall.  At this point, matter is still falling inward and accreting onto the PNS (see Fig.~\ref{fig:illustration}).

Soon after the bounce, the high-density core reaches approximate hydrostatic equilibrium, marking the birth of the PNS. The surface of this PNS is a source of intense neutrino emission.  The production of $\bar{\nu}_e$ and $\nu_e$ is enhanced over each of the $\nu_x$ flavors because accretion onto the PNS creates an extended mantle in which the $\nu_e$ and $\bar \nu_e$ neutrinospheres are at substantially larger radii than that of the $\nu_x$ flavors due to the reactions $e^- + p \leftrightarrow \nu_e + n$ and $e^+ + n \leftrightarrow \bar{\nu}_e + p$.  It is generally expected that neutrinos emitted from these outer layers of the PNS will deposit energy behind the shock and revive it.  Once the shock is revived, it will expel all the material outside, marking a successful explosion and leaving behind a cooling PNS.  This moment is shown in Fig.~\ref{fig:illustration} as a slight kink around 200~ms. In typical supernovae, the explosion time likely ranges between $\simeq$0.1--1~s, depending on the structure of the progenitor star and the efficacy of the neutrino mechanism.


\subsection{Post-explosion: PNS cooling}
\label{subsec:remnant}

After the shock wave is revived and a successful explosion ensues, the PNS enters a cooling phase through which it eventually becomes a NS.  The evolution of the PNS is characterized by a number of periods (see Fig.~\ref{fig:illustration}).  Over the first second of the cooling phase, the neutrino luminosity is driven by the cooling and contraction of the high-entropy, shock-heated outer layers---the mantle---of the PNS.  In reality, it is likely that while the explosion is occurring in some directions, mass accretion may continue in others~\cite[e.g][]{Lentz:15, Janka:16, Ott:18, Vartanyan:2019ssu, Burrows:20}.  The PNS decreases from its initial radius of $\simeq$50~km to close to its final radius of $\simeq$10~km, with the neutrino luminosity dropping substantially.

After cooling and contraction of the PNS mantle ends, the long-term PNS cooling phase begins (for details, see Ref.~\cite{Pons:99}).  Initially, the inner core of the PNS, which was not heated by the shock, is heated by the inward diffusion of neutrinos until the peak temperature of the PNS is at its center. The lepton number losses from the PNS drive the electron neutrino degeneracy parameter, $\eta_{\nu,e}$, toward zero (i.e., zero net  neutrino number). This is sometimes referred to as the  deleptonization period, although lepton number is lost throughout the entire cooling phase. The central temperature peaks and the central $\eta_{\nu,e}$ goes to zero about five to fifteen seconds after bounce, marking the end of this period. The duration of this period can depend strongly on whether or not convection occurs (see, e.g., Ref.~\cite{Roberts:12}) and on the size of in-medium corrections to the neutrino opacities (see, e.g., Ref.~\cite{Hudepohl:10}).

The final period is thermal cooling, in which neutrinos remove energy from the PNS over tens of seconds.  At this point, cooling is driven by temperature gradients alone, and no longer by gradients in the neutrino chemical potential. During this time, the dynamical changes are slow and the average energies of different flavor neutrinos have started to converge.  The time and spectral structures of the neutrino signal are strongly influenced by the properties of matter at and above nuclear density (see, e.g., Refs.~\cite{Keil:95, Pons:99, Roberts:12, Nakazato:2019ojk}) and it may be  possible to constrain the PNS radius based on observations of the neutrino  luminosities and spectra \cite{Hudepohl:10, Nakazato:2020ogl}.  This period ends when the PNS becomes optically thin to neutrinos, i.e., the onset of transparency, and the luminosity drops as the neutrinospheres recede. Transparency occurs as the matter cools and becomes more degenerate, decreasing neutrino energies and hence cross sections, and increasing final-state blocking of weak interactions.

During this transition from diffusion to optically thin cooling, the neutrino luminosity drops by over an order of magnitude over a few seconds as the neutrinospheres recede into the star.  Detecting neutrino transparency and being able to distinguish it from the signal of BH formation (see Fig.~\ref{fig:BH_formation_luminosity}) would provide important constraints on the properties of the PNS at high densities. The timescale of the cooling phase and the transparency time can be especially sensitive to the PNS mass and the properties of neutrino opacities in dense matter~\cite{Reddy:99, Burrows:98, Hudepohl:10, Roberts:12, Horowitz:16}.

Below, we frequently use the phrase ``late times" to indicate the part of the PNS cooling phase when the luminosities and average energies of the different flavors have sufficiently converged.  (The timescale for convergence is uncertain.  Importantly, it can begin before 10 s, which is the common prior understanding of ``late times," as marked in Fig.~\ref{fig:illustration}.)  Not only is PNS cooling the longest part of the neutrino signal, it is also the simplest to study and is the most under-explored phase.

A high-fidelity model of PNS cooling would be based on three-dimensional, multi-energy radiation hydrodynamic simulations of the PNS embedded in the post-explosion core-collapse environment.  Multi-dimensional hydrodynamics is important because it allows non-radial convective instabilities to develop inside the PNS.  Convectively driven fluid overturn can efficiently transport energy and lepton number through the PNS and accelerate neutrino cooling.  The very early phases of PNS cooling have been studied in axial symmetry and full 3D with multi-energy transport~\cite{Buras:06, Dessart:06, Radice:17, Nagakura:20}.  One calculation has been performed in axial symmetry out to very late time~\cite{Suwa:14}.  Simulations of core collapse  without imposed symmetries and multi-energy neutrino transport have only been performed for relatively short times ($\simeq$500~ms) due to their large computational cost (see, e.g., Refs.~\cite{Melson:15, Lentz:15, Roberts:16}). Therefore, most work on the late-time neutrino signal has relied on spherically symmetric simulations~\cite{Burrows:86, Wilson:88, Sumiyoshi:95, Keil:95, Pons:99, Fischer:10, Hudepohl:10, Roberts:12, Mirizzi:2015eza, Camelio:17}.  Although most of these simulations neglected convection, a subset~\cite{Wilson:88, Hudepohl:10, Roberts:12} has used a one-dimensional mixing-length theory of convection that can reasonably account for this hydrodynamic transport of energy and lepton number~\cite{Buras:06}.  In the PNS cooling phase, the only multi-dimensional effect expected to be important is convection, as other hydrodynamical instabilities, such as the standing accretion shock instability, occur outside of the PNS and will have been damped out by the explosion.


\subsection{Other possible late-time processes}
\label{sec:late_time}

There are additional mechanisms that may lead to late-time neutrino emission.  We discuss them only briefly because their fluxes are small and because our simulations do not include them.

Although the accretion rate onto the PNS should drop substantially after a successful supernova explosion, it will not go to zero. At early times ($t_{\rm{post-bounce}} \lesssim 5$ s), the asymmetry of the explosion itself may allow for continued accretion in cold streams along certain solid angles. This continued accretion is seen in successful three-dimensional explosion models (see, e.g., Refs.~\cite{Ott:18, Vartanyan:2019ssu, Burrows:20}). As this material falls back onto the PNS, a fraction of the released gravitational binding energy is converted into neutrino emission. This accretion-induced emission may compete with the diffusion luminosity from the PNS. 

Additionally, at even later times, the reverse shock produced by interaction of the SN shock with the density structure of the progenitor star can drive continued accretion or ``fallback''~\cite{1989ApJ...346..847C}. The exact history of this fallback depends on the detailed structure of the star~\cite{Zhang:2007nw}, but at late times the accretion tends toward steady flow free-fall accretion~\cite{1989ApJ...346..847C}, which gives an approximate contribution to the neutrino luminosity of
\begin{align}
L_{\nu,\text{fb}} \approx & \, 2.5 \times 10^{50} \, \text{erg} / \text{s} 
\left(\frac{M_\text{NS}}{1.4 M_\odot}\right) \left(\frac{M_\text{fb}}{0.01M_\odot}\right) \nonumber \\
& \left(\frac{R_\text{NS}}{10 \,\text{km}}\right)^{-1}
\times \left(\frac{t_0}{10 \,\text{s}}\right)^{-1}  \left(\frac{t}{t_0}\right)^{-5/3} ,
\end{align}
where $t_0$ is the time at which steady flow sets in ($\simeq$10~s gives a reasonable fit to the results of Ref.~\cite{Zhang:2007nw}). The fallback mass, $M_\text{fb}$, depends strongly on the progenitor structure, with more fallback in compact progenitors~\cite{Zhang:2007nw}.  A typical value for $M_\text{fb}$ is around 0.003 $M_\odot$~\cite{Sukhbold:2015wba}, and the corresponding $\bar{\nu}_e$ luminosity is shown in Fig.~\ref{fig:illustration}. In this case, the fallback flux dominates the PNS cooling flux due to the steep drop in the latter; the fallback flux appears linear due to the log-log scale.

Another possible source of late-time neutrino emission is nucleosynthesis.  Thermonuclear burning in shock heated material in the envelope of the star can produce radioactive isotopes, which subsequently undergo beta decay and emit neutrinos.  For comparison, the neutrino luminosity from nucleosynthesis in a Type Ia supernova is quite low, $\lesssim$ 10$^{49}$ erg/s for $\nu_e$, with a low average energy (and lower for $\bar{\nu}_e$)~\cite{Wright:2016xma, Wright:2016gar}.  Given that a typical core collapse produces $\sim 10\%$ as much nickel as a Ia~\cite{Sukhbold:2015wba}, this neutrino emission is negligible.


\subsection{Neutrino-mixing effects on emitted spectra}
\label{sec:mixing}

The astrophysical complexities of supernova neutrino emission are increased by the effects of neutrino mixing~\cite{Pantaleone:1992eq, Pantaleone:1992xh, Duan:2005cp, Sawyer:2005jk, Duan:2006an, Hannestad:2006nj, Fogli:2007bk, Sawyer:2008zs, Dasgupta:2009mg, Duan:2010bg, Mirizzi:2015eza, Chakraborty:2016yeg, Izaguirre:2016gsx, Capozzi:2018clo}.  The high density of matter changes the effective neutrino mixing parameters.  The high density of neutrinos also contributes to this, leading to complex collective phenomena, which might even cause equilibration of the spectra of all flavors through the so-called fast flavor conversion.  {\it Despite its importance, the neutrino mixing problem in core collapse is unsolved.}  Due to this uncertainty and the numerical complexity, neutrino-mixing effects are not included in supernova simulations.

Importantly, there is a factorization between neutrino emission and detection, i.e., detection depends on the supernova distance only through the overall $1/r^2$ dependence of the flux.  To address a common misconception, vacuum mixing en route from the supernova to Earth can be neglected.  In typical scenarios, neutrinos emerge in incoherent mass eigenstates due to the high matter densities they have passed through.  Even if not, phase averaging with realistic energy resolution would suppress any oscillatory terms on a distance scale small compared to the size of the supernova. This factorization allows use of the ``effective" (after mixing, just outside the supernova) flavor spectra for detection calculations instead of the initial (before mixing, as emitted) spectra.  Then, when a supernova happens, the data can be analyzed immediately, allowing comparisons between different detectors, different flavors, and to the SN 1987A data.  Before Sec.~\ref{sec:resultsNS}, we focus on the initial spectra, while in Sec.~\ref{sec:resultsNS}, we focus on the effective spectra.  It is a separate problem to relate the measured effective spectra to the initial spectra predicted from simulations to test neutrino mixing, as we discuss in Sec.~\ref{sec:physicsresults}.

A further simplification arises in the PNS cooling phase.  The flavor-dependent neutrinospheres converge over time and the temperature gradient in the emission region becomes more shallow.  Because of this, the luminosities and average energies of the different flavors should converge (and stay so at later times), as is the case for the simulation presented below.  We expect PNS cooling to be the dominant process at times beyond $\simeq$2~s or somewhat later.  Convergence of the spectra is expected to happen within several seconds, which will suppress the effects of active-flavor mixing, so the measured data will directly probe the astrophysics of PNS cooling.  This convergence must be tested with data, as we discuss.


\subsection{What neutrino data are needed}
\label{subsec:takeaway}

As is well known, it is important to measure well the early-time neutrino emission.  This will provide unprecedented probes of the dynamics of core collapse, accretion, and explosion.  It will also provide unprecedented probes of neutrino mixing in extreme environments.  However, disentangling all the physics effects will be challenging.

There are three reasons why it is important to measure well the late-time neutrino emission.  First, to measure the evolution of the PNS as it becomes a NS (or perhaps a BH).  Second, to measure the total radiated energy and lepton number, which depend significantly on the late-time emission.  Third, using the results of the previous two points, to help break degeneracies among all the contributing effects that shape the pre-explosion physics.

Success in this program requires measurements of the neutrino event rates to as late of times as possible, and certainly to at least a few tens of seconds to see the steep drop due to the onset of neutrino transparency.  But this is not enough.  Spectra must be measured to break the degeneracy in the detected event rate between number luminosity and average energy.  (We assume the core-collapse distance is reasonably known, as discussed in Appendix~\ref{sec:distance}.)  And all flavors must be measured to test if the number luminosities and average energies of the different flavors have converged.  The best channel to detect neutrinos of a given flavor needs to have a large cross section to ensure large yields, especially at late times.  It also needs to have clean kinematics to ensure that the neutrino spectrum can be estimated, at least statistically, from the measured energy-deposition spectrum. 

\begin{figure}[t]
\centering
\includegraphics[width=\columnwidth]{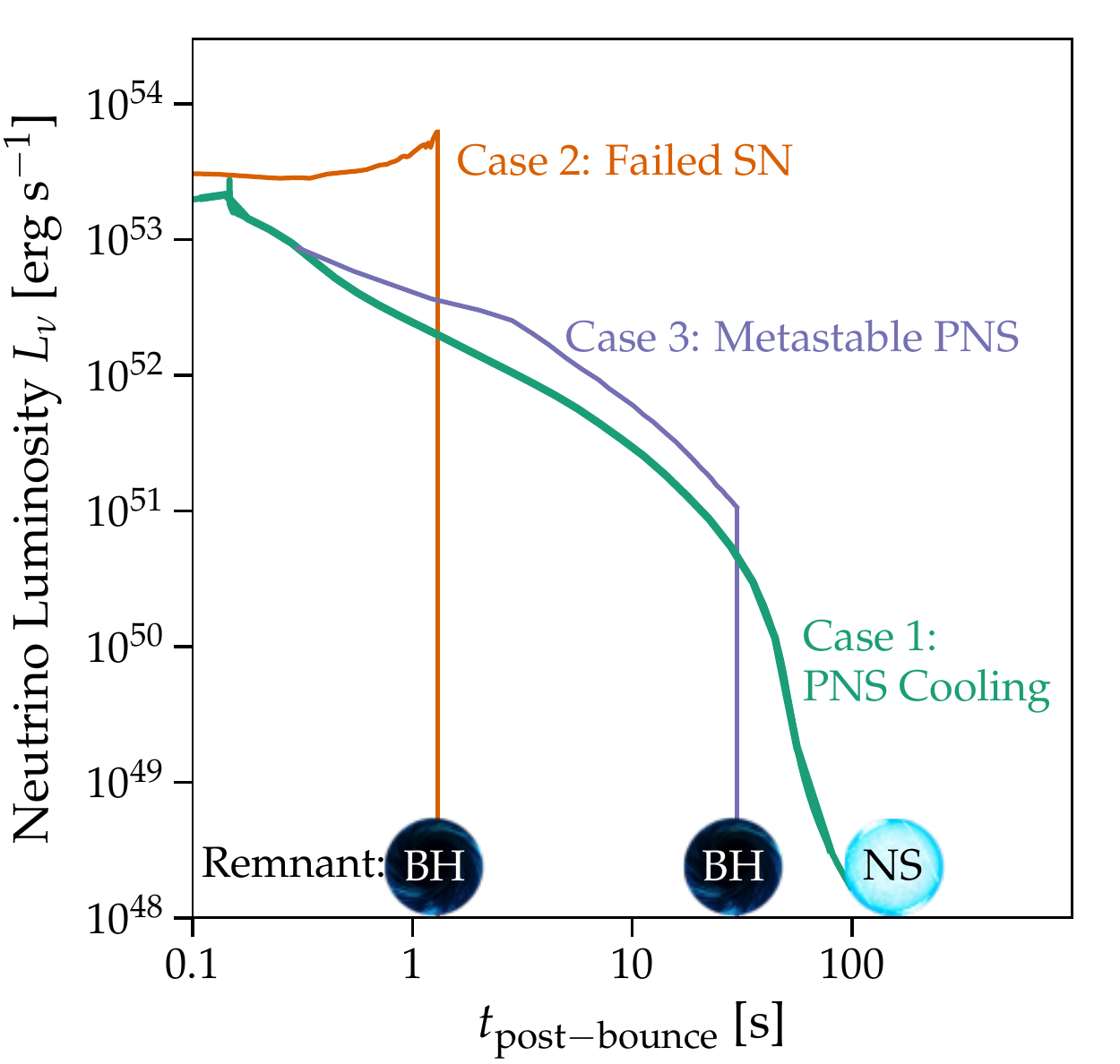}
\caption{Example total neutrino luminosities for three cases: PNS cooling, BH formation due to a failed supernova~\cite{Sumiyoshi:2007pp}, and BH formation due to a soft nuclear equation of state~\cite{Pons:01a}.}
\label{fig:BH_formation_luminosity}
\end{figure}


\subsection{Black hole formation}

In ordinary massive stars, core collapse begins with PNS formation.  However, this does not guarantee that the final state is a NS.  Below, we discuss two mechanisms by which a BH can form.  Compared to the NS-forming case, the neutrino luminosity before BH formation is larger, because the total energy release is larger (due to the larger, more compact remnant and increased accretion).  Once the BH forms, the flux is sharply truncated, with a sub-ms width, as discussed below.  How frequently BH formation occurs is not fully understood, and it depends on the progenitor mass, structure, and metallicity as well as the details of the SN explosion mechanism.  The fraction is thought to be between 5--50\%~\cite{Beacom:2000qy, Lien:2010yb, Horiuchi:2011zz, Sukhbold:2015wba, Woosley:2020mze}.

Figure~\ref{fig:BH_formation_luminosity} shows a model~\cite{Sumiyoshi:2007pp} with BH formation in the direct (or prompt) mechanism, labeled a ``Failed SN".   When the shock is stalled, material is accreting onto the hot PNS.  How much mass is accreted depends on the structure of the pre-collapse star and the time at which SN shock runaway begins.  If the PNS accretes enough mass to exceed the maximum allowed by the equation of state~\cite{Schneider:2020kxr}, it collapses into a BH. There is typically no successful explosion and no optical counterpart except for stellar disappearance~\cite{Kochanek:2008mp, Gerke:2014ooa, Adams:2016ffj}. 

If there is not a successful supernova explosion that ejects the envelope of the star, the time it takes for a BH to form varies with the compactness of the core, being shorter for more compact cores, which have a higher accretion rate~\cite{OConnor:2010moj}.  In Ref.~\cite{Schneider:2020kxr}, they calculate the time (post-bounce) for BH formation, assuming it occurs. (The mass ranges of progenitors that lead to BH formation are uncertain, and maybe even consist of many small islands~\cite{Sukhbold:2015wba}.) For the most compact cores they consider, with a progenitor mass of $\simeq$60 $M_\odot$, BH formation happens at $\simeq$0.9~s.  For the least compact core they consider, with a progenitor mass of 20 $M_\odot$, it happens at $\simeq$8~s~\cite{Schneider:2020kxr}.  For even smaller progenitors, down to $8 M_\odot$, scaling arguments, i.e., $t_{BH} \simeq 0.36\xi_{2.5}^{-1.64} \, \textrm{s}$~\cite{Schneider:2020kxr}, where $\xi_{2.5}$ is the compactness parameter~\cite{OConnor:2010moj}, suggest that if BH formation occurs (despite even spherically symmetric simulations of core collapse in these low-mass progenitors resulting in explosions~\cite{Hudepohl:10, Fischer:10}), then it could be as late as hundreds to thousands of seconds.  Some recent multi-dimensional simulations predict that low-mass progenitors in the range 12--15 $M_\odot$ will not produce successful explosions~\cite{Ott:18, Vartanyan:2019ssu, Burrows:20}; if they do fail, it would be within a minute.  In all cases, once the BH forms, the neutrino emission is sharply truncated and the entire star is quickly swallowed.

Figure~\ref{fig:BH_formation_luminosity} also shows a model~\cite{Pons:01a} with BH formation in the metastable (or delayed) mechanism, labeled a ``Metastable PNS" in Fig.~\ref{fig:BH_formation_luminosity}.  The luminosity is taken from a model that starts during the beginning of PNS cooling phase, at $\simeq$0.3~s after bounce.  The early-time luminosity should be the same as in Case 1 for the same progenitor.  Here the BH forms when a stable PNS becomes unstable due to a phase transition of the nuclear matter that softens the equation of state and hence lowers the maximum stable mass.  The phase transition may be triggered by the loss of lepton number due to neutrino emission, kaon condensation, or the creation of hyperons~\cite{Thorsson:1993bu, Keil:95, Glendenning1995, Prakash:1995uw, Prakash:97, Ellis:1996rp, Pons:99, Pons:01a}.  This would happen on the deleptonization timescale, around 10~s, and is thus allowed for SN 1987A. Generally, the presence of a phase transition at high density softens the high-density equation of state, which in turn reduces the maximum NS mass. Many of these earlier studies were performed before the discovery of NSs with masses $\gtrsim 2 M_\odot$ and used equations of state that could not support such massive NSs. It would be interesting to revisit these models in light of new constraints on the maximum NS mass.

In both cases, the duration of the truncation in the neutrino flux is very short.  In Ref.~\cite{Baumgarte:1996iu}, the authors ran a special singularity-avoiding dynamical simulation, finding the duration to be $\simeq$0.5~ms (the light-crossing time is $2R/c\simeq 0.1$~ms).  During this transition, the neutrino average energy plummets due to gravitational redshifting, making it even more difficult to measure the neutrinos.  Detecting neutrinos after the cutoff could indicate a propagation delay due to neutrino mass effects~\cite{Beacom:2000ng, Beacom:2000qy}.

BH formation can also occur on timescales longer than a minute or so post-bounce due to fallback (see Sec. \ref{sec:late_time}), where continued accretion onto an otherwise stable NS pushes it over the maximum mass.  Fallback-induced BH formation does not affect the neutrino signal, so we neglect it.


\section{Specific predictions for neutrino emission during PNS cooling}
\label{sec:theory} 

\begin{figure*}[t]
\centering	
\includegraphics[width=\textwidth]{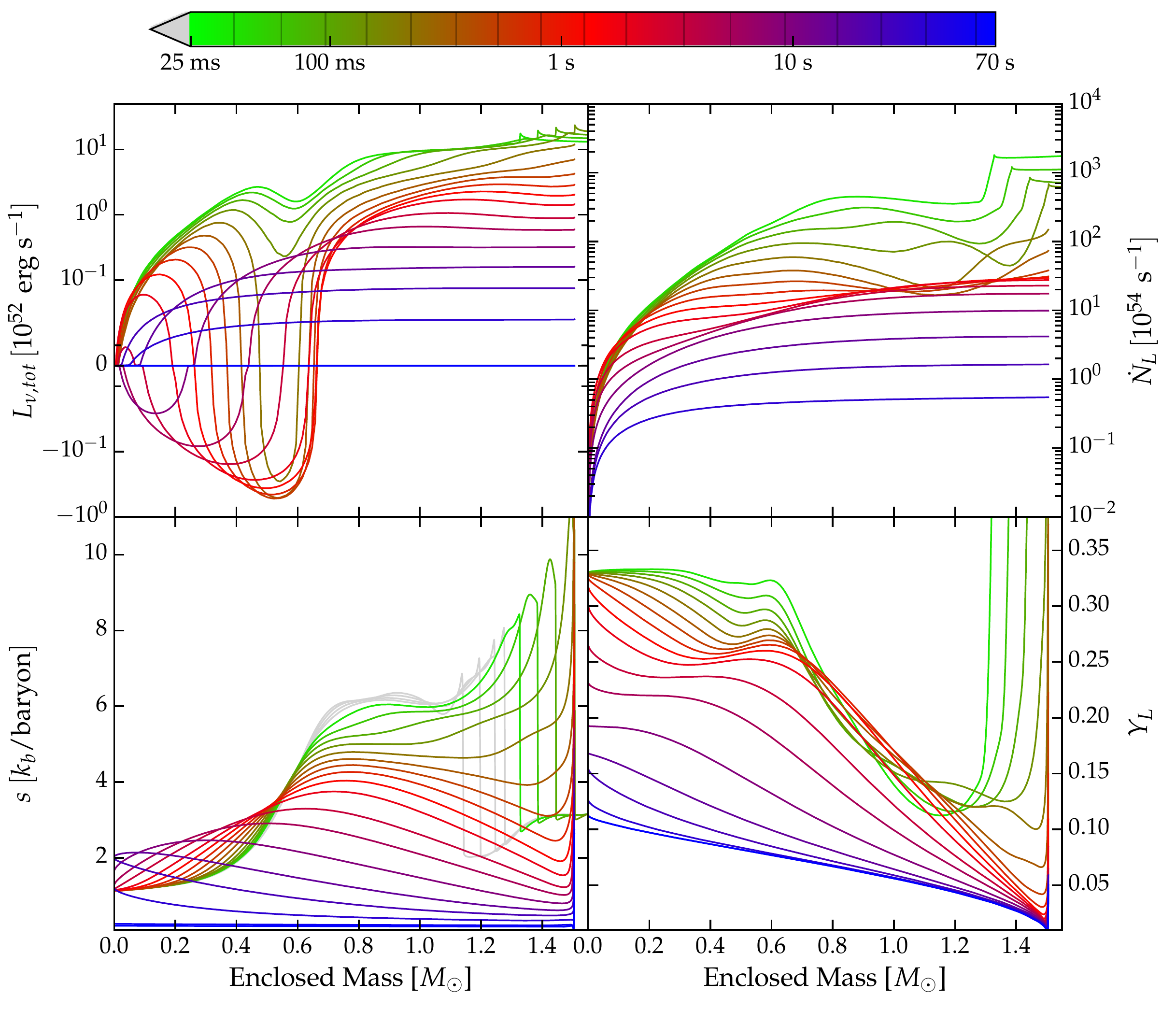}
\caption{Evolution of the internal structure of the PNS.  The panels show the total luminosity carried by neutrinos (upper left), total lepton number luminosity carried by neutrinos (upper right), entropy (lower left), and lepton fraction (lower right).  All quantities are plotted as a function of enclosed baryonic mass. The colors of the lines encode the post-bounce time, which can be read off the color bar at the top of the plot. The propagation of the SN shock is clearly visible until it is excised from the grid, as described in Sec.~\ref{sec:theory}.  After this, the cooling and deleptonization of the PNS occur on a timescale of tens of seconds.}
\label{fig:evolution}
\end{figure*}

In this section, we focus on the description of the PNS cooling phase for our nominal model, which guides our calculations of the late-time detection prospects.  The neutrino emission is mainly sensitive to the remnant mass and the lepton and entropy gradients left behind the shock.  These are both influenced to a greater or lesser extent by the physics of neutrino trapping, the details of the explosion mechanism, and the progenitor structure, but after these properties of the PNS are set the evolution is largely independent of other details of the progenitor structure. The neutrino emission profile could be materially altered from that in our nominal model by changes in the uncertain equation of state and neutrino opacities~\cite{Pons:00, Hudepohl:10, Roberts:12, Nakazato:2012qf, Suwa:2019svl, Nakazato:2019ojk, Nakazato:2020ogl}, as well as by the inclusion of fallback~\cite{1989ApJ...346..847C} or especially PNS convection~\cite{Roberts:12}.  However, while these could change the duration of the cooling phase (convection, in particular, would shorten it), we do not expect important qualitative changes in the neutrino emission profile.  As noted above, we do not include neutrino mixing in this section.


\subsection{Our simulation of PNS evolution}
\label{subsec:simulation}

\begin{figure*}[t]
\centering
\includegraphics[width=\columnwidth]{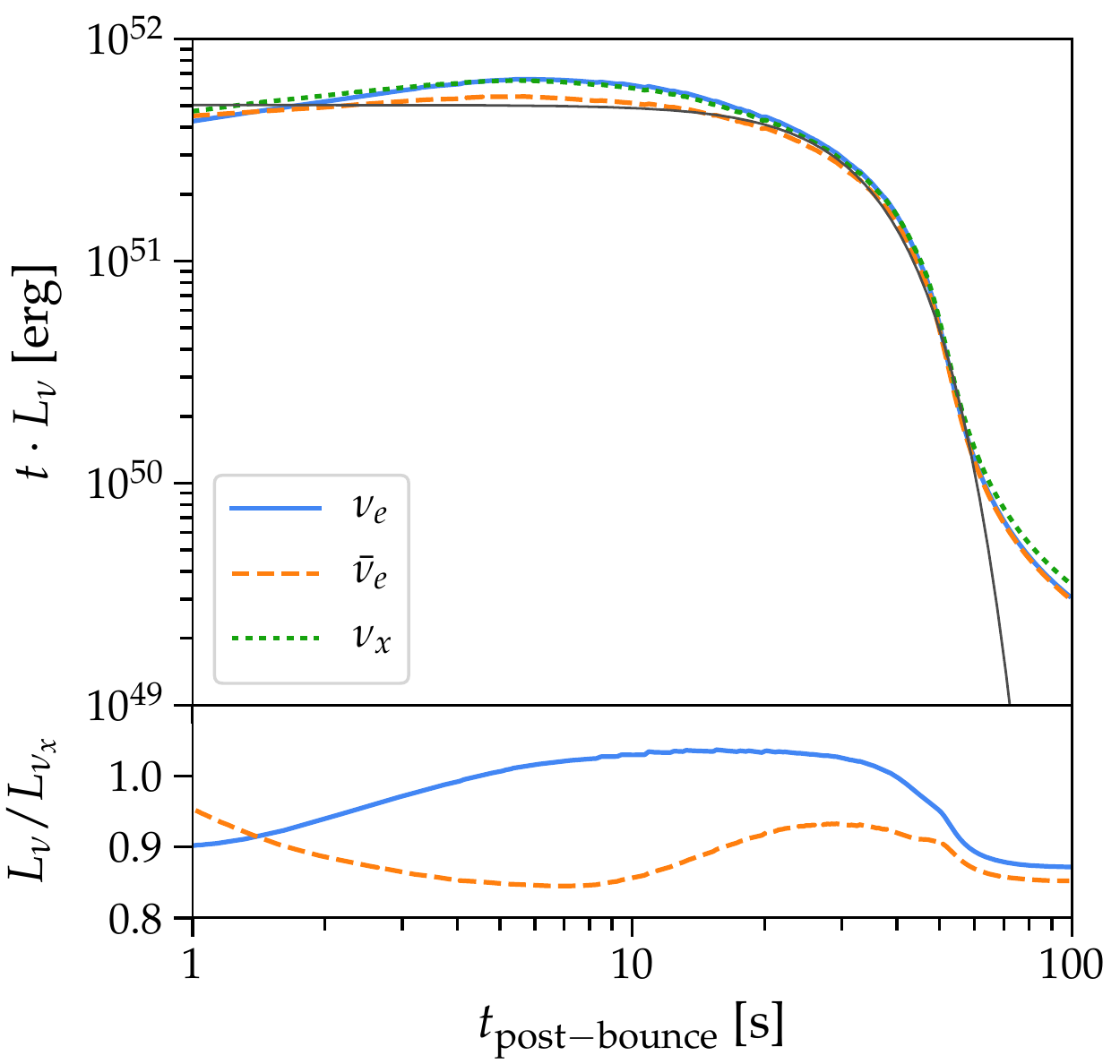}
\includegraphics[width=\columnwidth]{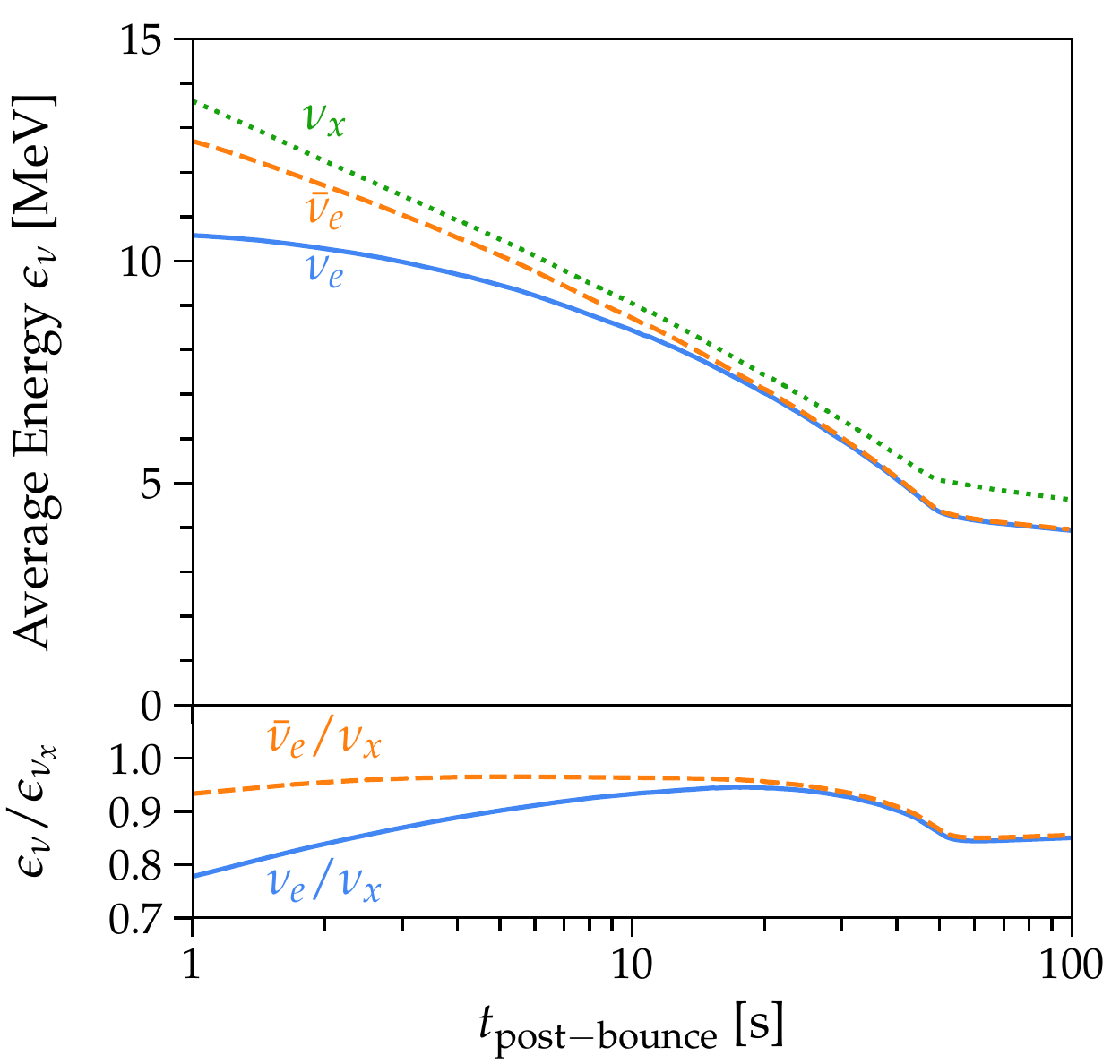}
\caption{Evolution of the luminosities (left panel) and average energies (right panel) of each flavor of neutrinos, as well as selected ratios.  The thin gray line is the functional fit in Eq.~\eqref{eq:luminosity_fit}.}
\label{fig:nominal_model}
\end{figure*}

We base our results on neutrino cooling curves produced by spherically symmetric neutrino radiation hydrodynamics simulations of pre- and post-core-collapse evolution of the PNS.  The calculations employ the two-moment neutrino transport method described in Ref.~\cite{Roberts:12a}, coupled to a spherically symmetric, implicit general relativistic hydrodynamics code based on the formalism of Ref.~\cite{Liebendorfer:01}.  We aim for a nominal model of the cooling with a relatively simple luminosity structure in time, so convection inside the PNS is ignored.  We employ the equation of state of Ref.~\cite{Schneider:17} using the Skyrme parameterization of Ref.~\cite{Lattimer:91}. We employ three-flavor neutrino transport, ignoring subtleties that lead to differences between $\nu_x$ and $\bar\nu_x$.  Neutrino interactions are calculated using the baseline rates of Ref.~\cite{Bruenn:85} with corrections to the charged-current reaction rates given in Refs.~\cite{Martinez-Pinedo:12, Roberts:12}.  We use bremsstrahlung rates from Ref.~\cite{Hannestad:98}.  Nuclear correlations can significantly reduce the neutrino opacity~\cite{Burrows:98, Reddy:99, Horowitz:2016gul, Burrows:2016ohd} and speed up cooling at late times~\cite{Reddy:99, Hudepohl:10}, although the corrections are only likely to be important once convection has ceased~\cite{Roberts:12}.  The size of these corrections is uncertain, due to lack of knowledge of the effective nuclear interaction at high density. We leave the investigation of the impact of different input choices on detection rates to future work.

In our nominal model of the neutrino signal, the simulation starts from the $15 \, M_\odot$ pre-collapse stellar model of Ref.~\cite{Woosley:02} (often denoted s15).  Collapse and deleptonization, core bounce, and accretion are simulated.  Once the SN shock passes a pre-chosen baryonic mass coordinate at $1.5 \, M_\odot$, the outer layers of the star are removed and replaced by a pressure boundary condition.  This is similar to the strategy that has been followed in many previous cooling calculations~\cite{Burrows:86, Keil:95, Pons:99, Roberts:12}, except that the ``initial'' conditions for the cooling are calculated self-consistently with the same code used to perform the cooling.  This can also be viewed as a method of inducing a supernova explosion, because it is well known that spherically symmetric models of core collapses do not explode~\cite{Liebendorfer:01a}, except for in a limited number of low mass progenitors~\cite{Fischer:10, Hudepohl:10}.  Nevertheless, the neutrino emission just after ``explosion'' is sensitive to the method by which the explosion is induced.  At later times, once the diffusion wave has reached deeper into the PNS, we expect the cooling to be fairly insensitive to the exact method by which the explosion occurred, assuming the baryonic mass of the PNS is constant. In the future, this needs to be verified with long-term, three-dimensional models of PNS cooling embedded in the core-collapse supernova environment, as discussed in Sec.~\ref{subsec:remnant}.  The cooling and deleptonization of the remnant object are simulated over approximately 100~s.


\subsection{Results from the nominal PNS cooling model}
\label{subsec:neutrino_emission}

Figure~\ref{fig:evolution} shows the evolution of key quantities in our nominal PNS cooling model, all in steps of post-bounce time.  The $x$-axis is the enclosed baryon mass, which is a Lagrangian coordinate that highlights the evolution of the entropy and lepton number of parcels of fluid as the PNS contracts.  The rapid entropy and net lepton number losses of the outer $0.4 \, M_\odot$ over the first second correspond to the phase of mantle cooling. During mantle contraction and early cooling, the central entropy of the PNS increases due to inward diffusion of neutrino energy (making the luminosity in Fig.~\ref{fig:evolution} negative) and Joule heating~\cite{Burrows:86, Pons:99}.  The core entropy peaks around fifteen seconds post-bounce, after which the entire PNS cools until it becomes optically thin.  The evolution of the neutrino diffusion wave and the deleptonization of the core are also clearly visible.  For a more detailed description of the interior evolution of the PNS, including the evolution of the density and temperature, see Ref.~\cite{Roberts:17}.

\begin{figure*}[t]
\centering
\includegraphics[width=\columnwidth]{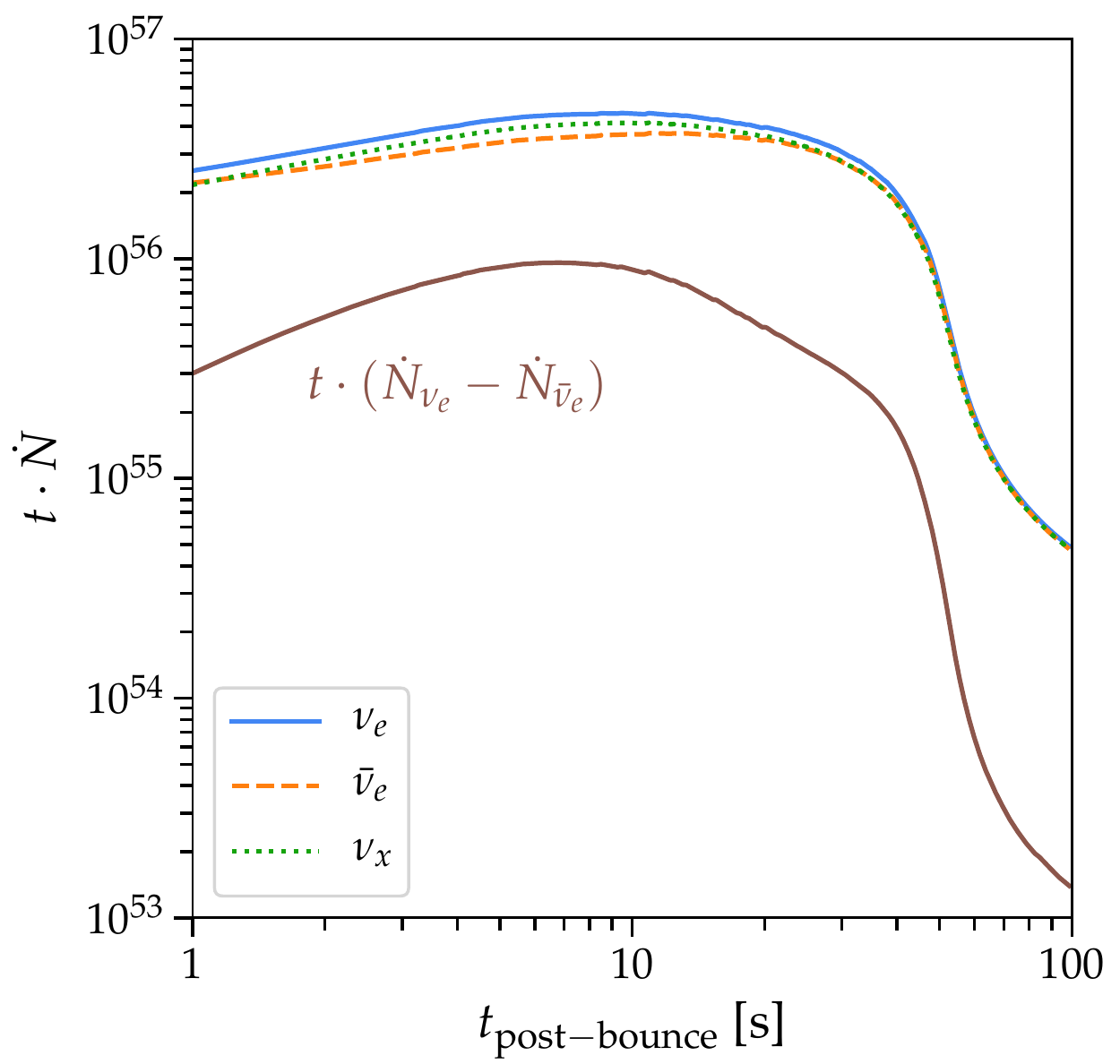}
\includegraphics[width=\columnwidth]{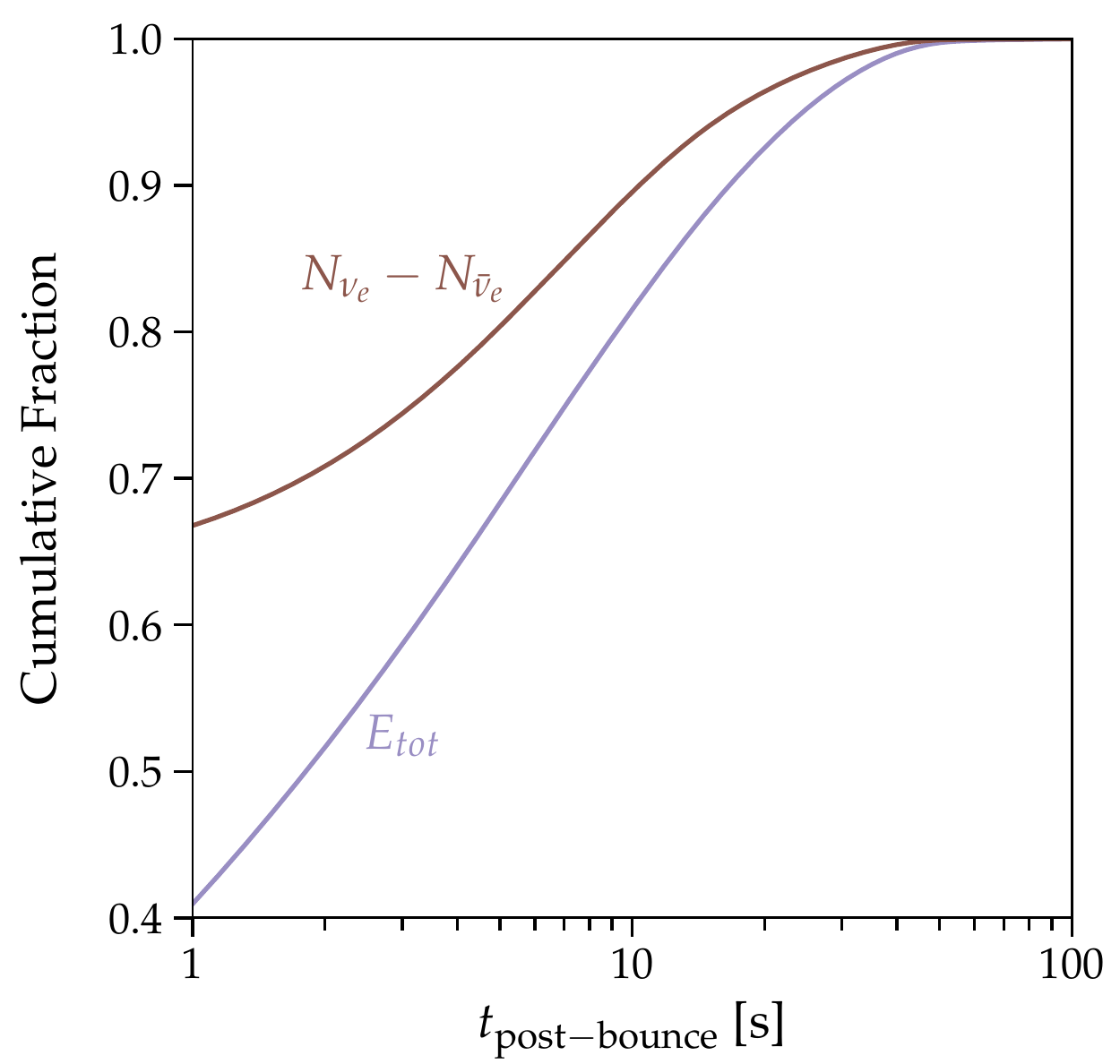}
\caption{{\bf Left:} Evolution of the number fluxes and the electron-flavors difference. 
{\bf Right:} Cumulative fractions of total emitted energy and lepton number (note that the $x$-axis does not start at zero).}
\label{fig:nominal_number} 
\end{figure*}

Figure~\ref{fig:nominal_model} (left panel) shows the evolution of the neutrino luminosities in our nominal model.  Past one second, those of the different flavors are within 10\% of each other.  The luminosities can be reasonably fit with the form
\begin{equation}
\label{eq:luminosity_fit}
L \propto t^{-1} e^{-(t/\tau_c)^\alpha},
\end{equation}
where $\tau_c$ is a characteristic timescale for the cooling, the transparency time, with $\tau_c = 36$~s and $\alpha = 2.66$.  By the time of the onset of transparency, the quantity $t \, L$ has fallen off by roughly a factor of 10, so that $L$ has fallen off much more.

As a general principle, when considering the evolution of a quantity $dA/dt$, if one uses a logarithmic $x$-axis, then one should plot $t \, dA/dt = dA/d\ln t = (2.3)^{-1} \, dA/d\log{t}$, so that the differential in the derivative on the $y$-axis matches the increment on the $x$-axis.  This ensures that the relative heights of the curve at different times accurately reflect their relative contributions to the integral $\int dt \, (dA/dt) = \int d\ln t \, (t \, dA/dt)$.  Accordingly, we do this for Fig.~\ref{fig:nominal_model} and subsequent figures.  In the specific case considered here, this factor of $t$ happens to cancel the $1/t$ in Eq.~(\ref{eq:luminosity_fit}), which indicates that the energy loss per log-time interval is constant until the exponential becomes important.  This suggests a simple underlying physical principle, but we have not identified one.

Physically, the luminosity emerging from the PNS is given by $L_\nu = 4 \pi \phi_\nu R_{\nu}^2 T_{\nu}^4$, where $R_{\nu}$ is the average radius of neutrino decoupling (also called the neutrinosphere), $T_{\nu}$ ($\simeq$3$E_\nu$) is the temperature at $R_\nu$, and $\phi_\nu$ is a dimensionless correction factor that accounts for deviation from black-body emission~\cite{Hudepohl:10}.  This expression is useful for a rough understanding of the neutrino signal, but many details of the neutrino emission are subsumed into $\phi_\nu$.  In the first twenty seconds, the luminosity follows a $1/t$ behavior due to the combined evolution of $T_{\nu}$ and $R_{\nu}$, the latter of which stays close to the PNS radius, although the time dependence of $T_{\nu}$ and $R_{\nu}$ do not follow simple analytic forms (see Fig.~3 in Ref.~\cite{Roberts:17}). Similar time dependence over the whole PNS evolution is observed in the cooling calculations of Ref.~\cite{Pons:99}, as pointed out by Ref.~\cite{Metzger:07}. Analytic models that assume a constant PNS radius (and $R_\nu$) recover a different functional form of the cooling timescale~\cite{Prakash:97, Roberts:16}.

Figure~\ref{fig:nominal_model} (right panel) shows the evolution of the average neutrino energies.  Compared to the luminosities, here the differences between flavors are slightly larger.  At early times, the $\nu_e$ opacity is higher in the outer layers of the PNS due to the large neutron density, so the $\nu_e$ decouple near the surface of the PNS.  The $\bar{\nu}_e$ decouple at a smaller radius, because their charged-current interactions are with protons, which are subdominant.  The $\nu_x$ decouple at an even smaller radius because they experience no charged-current interactions, due to the large masses of the corresponding leptons.  Because the temperature increasing with decreasing radius in the PNS, $\nu_x$ have the highest average energies, the $\bar{\nu}_e$ have the second-highest, and the $\nu_e$ have the lowest.  As time goes on, the temperature profile of the outer layers of the PNS flattens out and the neutrinospheres converge, which causes the late-time convergence of the average neutrino energies and luminosities~\cite{Fischer:11}. In the PNS cooling model used here, even at the latest times, $\nu_x$ has slightly higher average energy than $\nu_e$ and $\bar\nu_e$. This difference is likely due to the implementation of neutrino bremsstrahlung used in the simulation, which assumes a thermal distribution for one of the pair of neutrinos produced, be we are not certain.  Because the difference is within our model uncertainties, we neglect it here but will investigate it further in future work.  In some other models, all flavors converge at late times~\cite{Hudepohl:10, Fischer:10}. Accordingly, we neglect neutrino-mixing effects.

Figure~\ref{fig:nominal_number} (left panel) shows the neutrino number fluxes, which are luminosities over average energies.  It is not surprising that number fluxes also roughly follow a $1/t$ trend in the first twenty seconds and then decrease sharply when the PNS becomes transparent.  Figure~\ref{fig:nominal_number} (left panel) also shows the lepton number loss, which drops faster than the luminosity after about ten seconds, corresponding to  the end of the deleptonization period.

Figure~\ref{fig:nominal_number} (right panel) shows the cumulative fractions of energy and lepton number, which reveals how different time periods contribute to the time-integrated emission.  In our model, the total neutrino luminosity (integrated from $t=0$~s) is $2\times10^{53}$~erg and the total lepton number loss is $7 \times 10^{56}$.  About 60\% of the energy is emitted during the cooling phase.  Measuring the emission to late times is critical to measuring the total energy and lepton-number loss, both to probe core collapse and to test for new physics (e.g., Refs.~\cite{Raffelt:96, Fischer:2016cyd, Chang:2018rso, Suliga:2020vpz}).


\subsection{Model uncertainties}
\label{subsec:variation}

The structure and timescale of PNS emission can be influenced by a number of physics inputs that are uncertain or not included in our nominal model.  First, the initial conditions of the PNS (i.e., its total mass and the initial distributions of lepton number and entropy) will set the total energy and lepton number available during the cooling epoch and determine the structure of the PNS.  The PNS mass will certainly vary with progenitor star, which will in turn change the timescale of PNS emission~\cite{Pons:99}.  Second, convection enhances the early PNS cooling neutrino luminosity and deleptonization, shortens the duration of the neutrino signal, and introduces new features in the neutrino signal~\cite{Burrows:87, Roberts:12, Mirizzi:2015eza}.  Third, the uncertain equation of state and neutrino opacities of dense matter can alter the neutrino diffusion rate and change the time structure of the late-time neutrino signal~\cite{Burrows:98, Reddy:99, Hudepohl:10, Roberts:12, Horowitz:16, Nakazato:2018, Nakazato:2019ojk, Nakazato:2020ogl}.  Finally, processes beyond the standard model, such as axion production, may provide another channel for energy loss from the PNS and impact the timescale of neutrino emission~\cite{Raffelt:96, Fischer:2016cyd, Chang:2018rso, Suliga:2020vpz}.  Nevertheless, the rapid drop in the neutrino signal at the onset of transparency seen in our nominal model is likely to be generic, though the transparency time may vary.

In future work, we will explore in detail how different micro-physics choices affect the late-time neutrino signals and how we can disentangle them.  This will build on the work in Refs.~\cite{Nakazato:2012qf, Suwa:2019svl, Nakazato:2019ojk, Nakazato:2020ogl}, which consider a wide range of inputs, especially the equation of state.


\section{Detection Prospects---Neutron-Star Case}
\label{sec:resultsNS}

In this section, we calculate the detection prospects for all flavors, both event rates and spectra, for the case where a NS is formed.  We assume that the supernova is detected in electromagnetic bands and that the neutrino signal is well measured at early times, so that the supernova direction and distance are reasonably known (see Appendix~\ref{sec:distance}).  In the following, we review general points about detection (Sec.~\ref{subsec:detection}), and then present our calculations of the measurement prospects for $\bar{\nu}_e$ (Sec.~\ref{subsec:SK}), $\nu_e$ (Sec.~\ref{subsec:DUNE}), and $\nu_x$ (Sec.~\ref{subsec:JUNO}).

For a real supernova, the measurements discussed in this section would be of the effective flavor spectra (after any neutrino mixing effects), and we consider them as such.  As a reasonable proxy, because our focus is on developing the overall calculational picture and the projected uncertainties, as a numerical input we use the initial flavor spectra predicted by the simulation in Sec.~\ref{sec:theory}.  In Sec.~\ref{sec:physicsresults}, we discuss mixing effects further.


\subsection{Core-collapse neutrino detection}
\label{subsec:detection}

We calculate the neutrino signals in three large detectors---Super-K, DUNE, and JUNO---as described in detail below.  For neutrinos of a given flavor, the event rate spectrum in terms of detected energy, $E_{det}$, is
\begin{equation}
\frac{dR}{dE_{det}}(t) =
\frac{N_\mathrm{target}}{4\pi D^2}
\frac{L(t)}{\langle E \rangle(t)}
\int dE_\nu \, f(E_\nu, t) \, \frac{d\sigma}{dE_{det}}(E_\nu, E_{det}) \,,
\end{equation}
where $L$ is the luminosity and $\langle E \rangle$ the average neutrino energy (so that their ratio is the rate of emitted neutrinos), $f(E_\nu)$ is the normalized neutrino spectrum, and $d\sigma$ is the differential cross section. The kinematics relating $E_\nu$ to $E_{det}$ depend on the interaction.  To compare with observables, the calculated $E_{det}$ spectrum must be further convolved with detector energy response and efficiency.  For simplicity above, we omit the notation to describe those detector-dependent effects, but we include them in our calculations, as described below.  Backgrounds, which are  important, are also described below.

We omit discussion of the many other detectors that would be sensitive to the main signal of a Milky Way core collapse but which would have low yields at late times~\cite{Scholberg:2012id}.  IceCube and KM3NeT will have huge yields, but a core-collapse signal can only be detected through excesses over the large background rates, and spectra cannot be measured~\cite{Kopke:2017req,  ColomerMolla:2020qds}.  A recent IceCube thesis suggests that the supernova signal can be detected over background to about 15 s~\cite{Eberhardt:2017vce}. With new types of photosensors, these neutrino observatories might be able to detect individual events and crude measures of their energy~\cite{Boser:2013oaa}, which could enable very late-time measurements. These points should be explored.

\begin{figure*}
\centering
\includegraphics[width=\columnwidth]{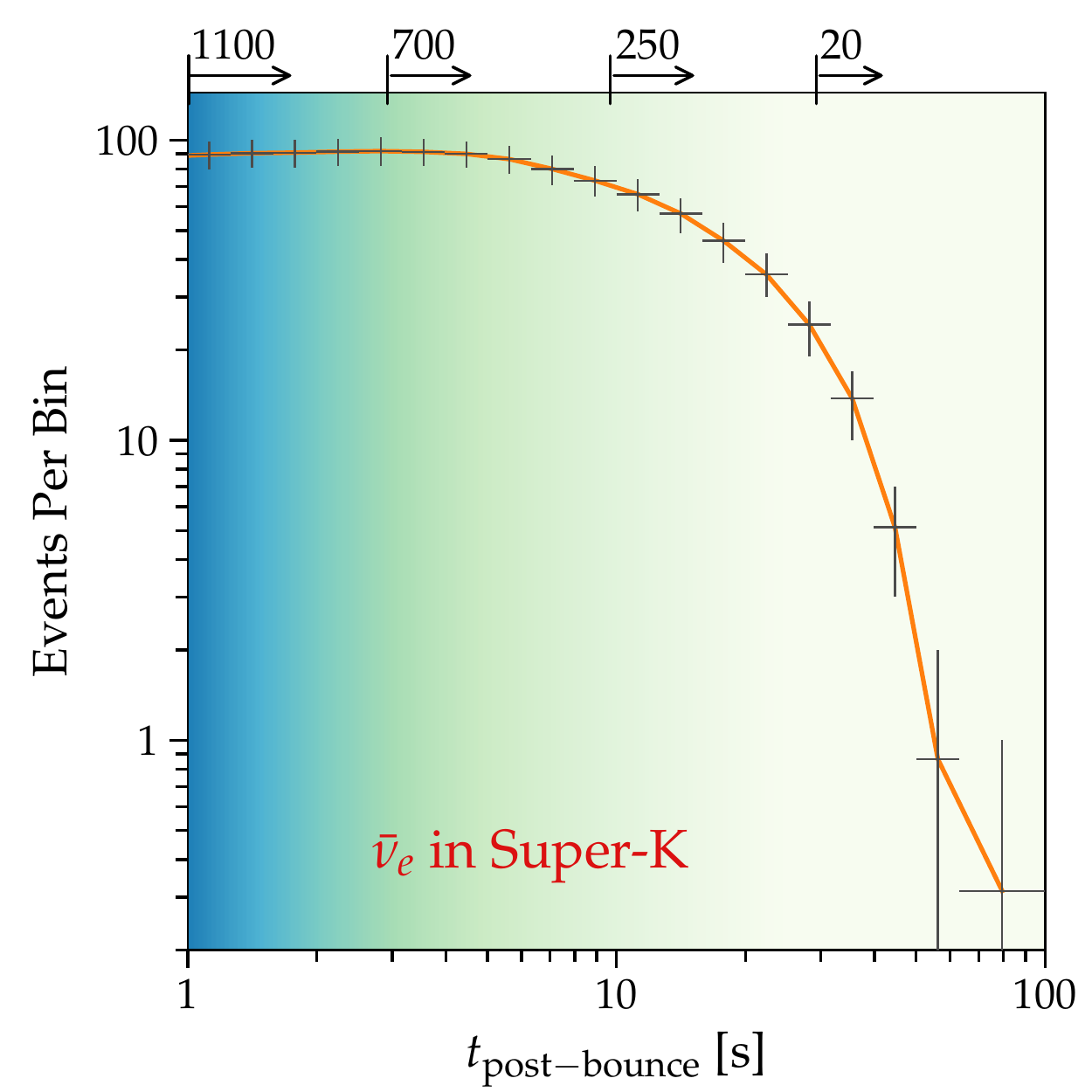}
\includegraphics[width=\columnwidth]{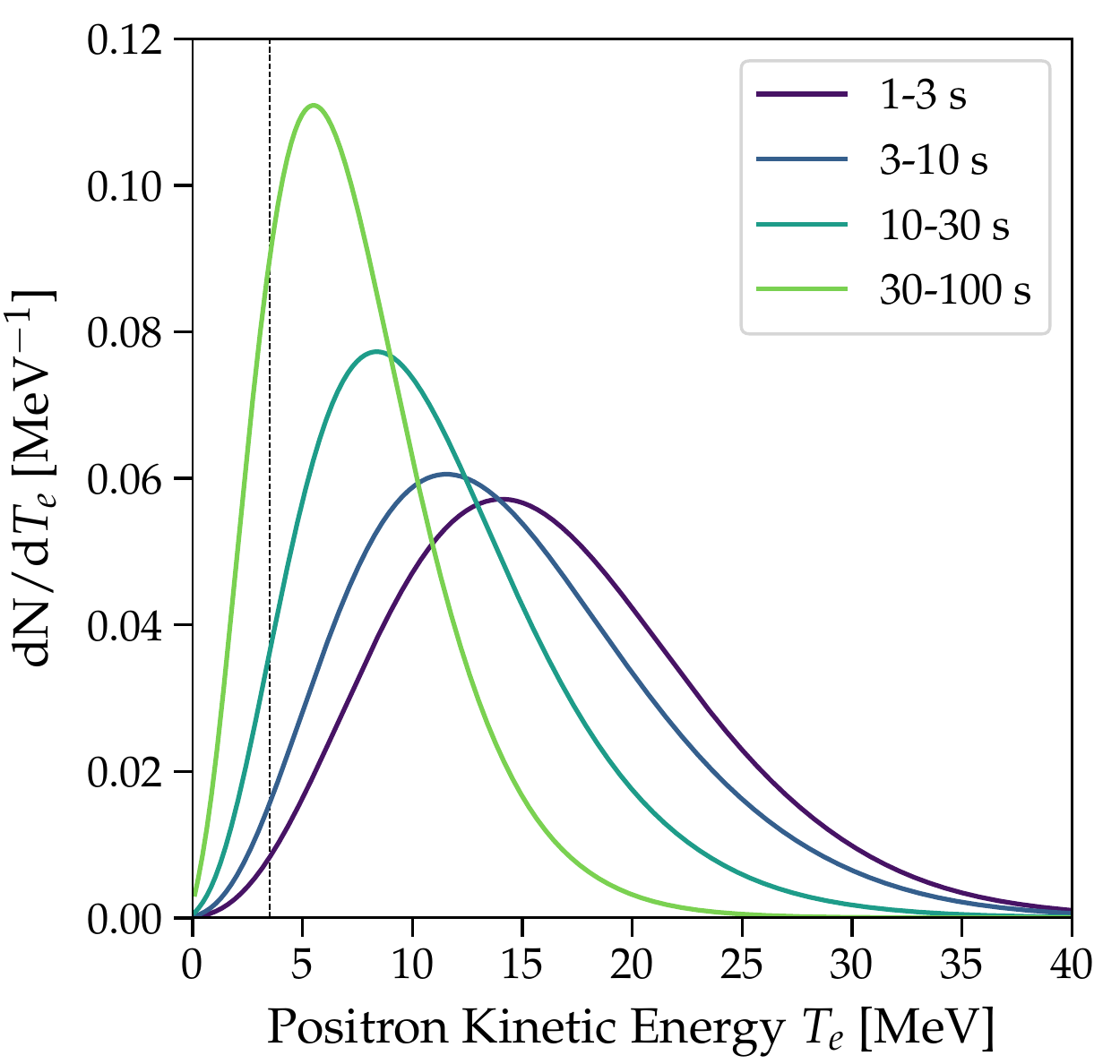}
\caption{{\bf Left:} Event rate (more precisely, the counts in each log-time bin) of $\bar{\nu}_e$ in Super-K due to the inverse-beta interaction (with a detector threshold of 3.5~MeV).  The vertical error bars show the Poisson uncertainties on the counts and the horizontal error bars show the bin widths.  Cumulative counts remaining beyond selected times (1, 3, 10, and 30 s) are shown on the top axis.  At late times (green shading), PNS cooling is the dominant emission mechanism and our PNS model is at its most reliable.  At early times (blue shading, with an uncertain boundary), these assumptions are less valid.
{\bf Right:} Spectra of positron kinetic energy in Super-K in selected time ranges, as marked.  The spectra are individually normalized to facilitate comparison of their shapes.  The vertical dotted line at 3.5 MeV indicates the assumed Super-K threshold.}
\label{fig:spectrum_SK}
\end{figure*}

Throughout, we assume that the core collapse is at a distance $D = 10$~kpc.  This is somewhat larger than the expected average distance (see Appendix~\ref{sec:distance}), but it is conventional and allows easy scaling.  The vast majority of Milky Way core collapses are within 5--15 kpc (with very few beyond 20 kpc)~\cite{Adams:2013ana}, so that the yields could be at worst a few times smaller than we assume, with the qualitative picture unchanged.  

Though the probability of the core collapse being extremely close (0.1--1 kpc) is low, it is possible, and the higher counts would make all of our conclusions stronger.  This is especially true for $\nu_x$ detection in JUNO, where detector backgrounds are important for a collapse at 10~kpc, but which would be negligible for a much closer event.  In addition, for a nearby burst, it may become possible to detect the fallback flux, which could be distinguished by a rise in the ratio of $\nu_e + \bar{\nu}_e$ relative to $\nu_x$.  A general concern is if the data-acquisition systems can accommodate such large rate without loss; this is outside the scope of our work, but the experiments should give it serious attention.

\subsection{$\bar{\nu}_e$ in Super-K}
\label{subsec:SK}

For $\bar{\nu}_e$, the current best detector is Super-K~\cite{Abe:2016nxk}, using the inverse-beta interaction with free (hydrogen) protons,
\begin{equation}
\bar{\nu}_e + p \rightarrow e^+ + n. 
\end{equation}
JUNO and Hyper-K will also have excellent capabilities for this channel, as discussed below. We neglect neutrino-electron scattering events because of the small yields and because all flavors contribute, and we are focusing on the cleanest measurements of each flavor.

Super-K's inner detector is a tank of 32 kton of ultrapure water~\cite{Abe:2016nxk} with gadolinium being added now.  Relativistic charged particles produce Cherenkov light that is viewed by photomultiplier tubes mounted on the surface.  To be conservative about the prospects for detecting the latest-time events, we assume a detector mass of 22.5 kton, corresponding to the fiducial volume used for solar-neutrino studies, where the signal is a single electron, and backgrounds must be minimized.  Nearly all core-collapse neutrino studies assume 32 kton, which is conventional but not fully realistic.  For electrons (or positrons, which cannot be distinguished), the detection threshold for solar-neutrino studies is 3.5~MeV kinetic energy.  We apply energy smearing following Ref.~\cite{Abe:2016nxk} and take the detection efficiency to be perfect above this energy and zero below.  For solar-neutrino studies, the efficiency is somewhat lower due to the strong cuts needed to reject backgrounds, but during a core-collapse burst, much milder cuts can be used due to the high signal rate.

The presence of neutrons can be detected through their captures after thermalization, which give a coincidence signal delayed from the positrons produced in inverse beta decay.  In water, $\simeq$20\% of these neutrons can be detected through their radiative captures on free protons, which produce 2.2-MeV gamma rays~\cite{SK:2015xra}.  In Super-K with dissolved gadolinium (the GADZOOKS! proposal of Beacom and Vagins~\cite{Beacom:2003nk}), radiative captures on gadolinium lead to an $\simeq$8-MeV gamma-ray cascade, which is easy to detect.  Gadolinium is now being added to Super-K~\cite{Ikeda:2019pcm}, and ultimately a neutron-tagging efficiency of $\simeq$90\% is expected.  We assume that inverse-beta events are detected as coincident pairs; even without gadolinium, event separation will be good, as discussed below.

For the inverse-beta interactions, the reaction threshold is $E_\nu = 1.806$ MeV and the cross section is precisely known~\cite{Vogel:1999zy, Strumia:2003zx}.  The kinematics are favorable, with $T_e \simeq E_\nu - 1.8$~MeV, and the angular distribution is nearly isotropic.  We evaluate the cross section and kinematics including the neutron-recoil and weak-magnetism corrections~\cite{Vogel:1999zy}.  The neutrino energy can be inferred, event by event, from the detected positron kinetic energy.  This allows the best measurement of the time evolution of the neutrino average energy.  Detecting neutrons does not affect neutrino energy reconstruction, but it will greatly reduce any backgrounds.

Figure~\ref{fig:spectrum_SK} (left panel) shows the event rate of inverse-beta interactions in Super-K for our nominal model.  We adopt a log scale for the $x$-axis, choosing 10 equal-width bins per factor of ten in time.  For the number of counts in a bin, we plot $\Delta N = 2.3 \, (t \, dN/dt) \, \Delta \log_{10}{t}$, as per the discussion above.  Due to this factor $t$, the drop in the count rate $dN/dt$ is even more dramatic than it appears.

The most important result is that there are many events during the PNS cooling phase.  The cumulative counts beyond selected times are noted on the top axis.  After $\simeq$10 s, the end of the SN 1987A signal, Super-K would expect 250 events.  The shape of the curve is similar to that of the luminosity in Fig.~\ref{fig:nominal_model}.  The correspondence is not exact because the neutrino spectrum changes with time.  The last event, defined by where the reverse cumulative event count crosses one, would be detected at $\simeq$50~s, late enough to robustly measure the entire transition to transparency, which is truly remarkable.

Even without gadolinium, the inverse-beta events can be isolated from other events with high purity. The total yields from other core-collapse neutrino interactions are below 10\%~\cite{Ikeda:2007sa, Abe:2016waf}; those on oxygen~\cite{Haxton:1987kc, Kolbe:2002gk, Nakazato:2018xkv} are suppressed by the low average neutrino energies and those on electrons can be cut by avoiding a small solid angle in the forward direction.  Detector backgrounds in Super-K are negligible during the neutrino burst, even at late times~\cite{Abe:2016nxk}.  In Ref.~\cite{Pons:01a}, an early work on late-time emission, the authors assumed a very high Super-K detector background rate of $\simeq$12~min$^{-1}$ (the energy threshold was unspecified), which cut off their neutrino signal at $\simeq$30~s.  Super-K has made excellent progress on reducing detector backgrounds~\cite{Bays:2011si, SK:2015xra, Nakano:2019bnr}.  The largest backgrounds are from intrinsic radioactivities (dominant below about 6 MeV) and muon-induced radioactivities (dominant above).  The total background rate above 3.5~MeV after standard solar-neutrino cuts is only $\simeq$0.21~min$^{-1}$, with 0.16 from radioactivity and 0.05 from spallation~\cite{Abe:2016nxk}.  Solar neutrino events have a rate of $\simeq$0.013 min$^{-1}$.  Backgrounds will be suppressed further by the addition of gadolinium, as the signal has a neutron, while most detector backgrounds and other core-collapse neutrino interactions do not.

Figure~\ref{fig:spectrum_SK} (right panel) shows the spectra of inverse-beta interactions in Super-K for our nominal model.  Prior works have presented only the event rate, ignoring the spectrum.  In fact, a key test of the PNS cooling model is how the average energy varies with time.  For $\bar{\nu}_e$, it will be straightforward to measure the spectrum and its time evolution.  Key reasons are the tight relationship between positron energy and neutrino energy, the good energy resolution, and the high statistics.  Within the first 10 seconds, the spectra peak between 10--15 MeV, well above threshold.  At later times, as the average energy falls (see Fig.~\ref{fig:nominal_model}), the peak moves to lower energies, but remains above threshold, which is critical to an accurate measurement.  As we show below, measuring spectra for the other flavors will be more difficult.  In future work, we will perform detailed studies of how to reconstruct the neutrino spectra and develop the implications.

For the JUNO detector~\cite{An:2015jdp}, the $\bar \nu_e$ event yield would be comparable, and the energy resolution and threshold even better.  Due to a shallower depth, JUNO's spallation background rates will be higher, but techniques to reduce the backgrounds seem adequate~\cite{Li:2014sea, Li:2015kpa, Li:2015lxa, Zhao:2016brs, Abusleme:2020zyc}.  For the Hyper-K detector, which is $\simeq$8 times bigger than Super-K, the results would be correspondingly better, though it will also be at a shallower depth~\cite{Abe:2018uyc}, and new techniques to reduce spallation backgrounds~\cite{Li:2014sea, Li:2015kpa, Li:2015lxa} will be important for realizing its full potential.  In addition, Hyper-K's nominal design uses pure water, which makes background rejection less efficient.  We strongly encourage continued consideration of adding gadolinium to Hyper-K.

In summary for the Super-K measurement of $\bar{\nu}_e$, it will be near-ideal, allowing clean measurements of the spectrum to late times and of the count rate to very late times, providing precise tests of PNS emission models.  However, one flavor is not enough to prove that the neutrino emission is dominantly from PNS cooling, so the other flavors must also be probed, as we discuss next.


\subsection{$\nu_e$ in DUNE}
\label{subsec:DUNE}

\begin{figure*}[t]
\centering
\includegraphics[width=\columnwidth]{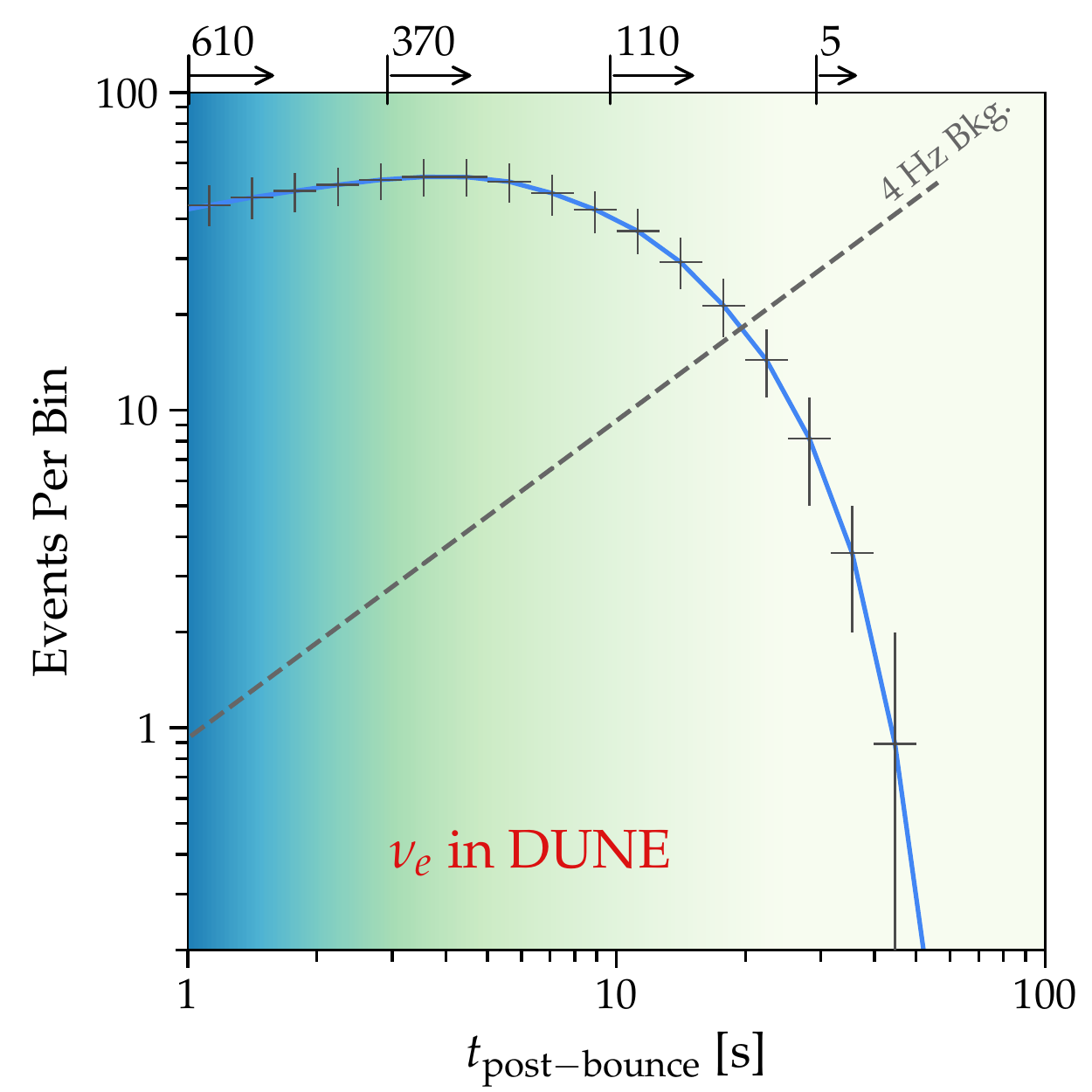}
\includegraphics[width=\columnwidth]{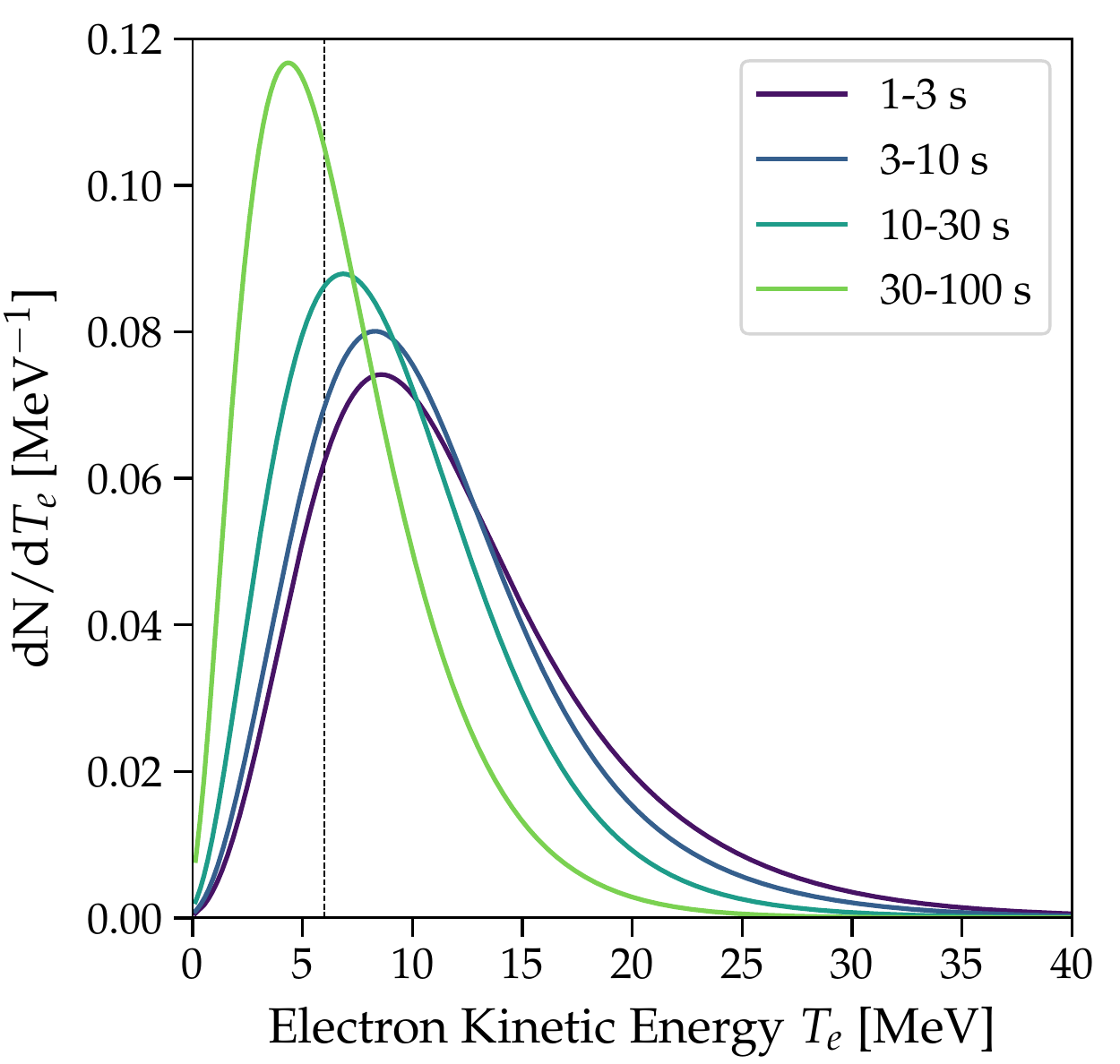}
\caption{Similar to Fig.~\ref{fig:spectrum_SK}, but for $\nu_e$ events in DUNE due to neutrino-argon interactions (first case, based on primary-electron energy).
{\bf Left:} Event rate for a threshold of 6~MeV in electron kinetic energy.  In addition to the signal rate, we show the background rate using the same bins (equal steps in $\log_{10}{t}$).  Because the background rate $dN/dt$ is constant, the number of background counts per bin, $\Delta N = 2.3 \, (t \, dN/dt) \, \Delta \log_{10}{t}$, rises linearly with $t$.
{\bf Right:} For the spectra, the vertical dotted line at 6 MeV indicates the assumed DUNE threshold.}
\label{fig:spectrum_DUNE}
\end{figure*}

For $\nu_e$, the best detector will be DUNE~\cite{Abi:2020evt}, using the neutrino-argon interaction
\begin{equation}
\nu_e + \, ^{40}\mathrm{Ar} \rightarrow e^- + \, ^{40}\mathrm{K}^*,
\end{equation}
where the star denotes a nuclear excited state.  This channel is much more favorable than $\nu_e$ interactions with carbon in JUNO or with oxygen in Super-K because these have small yields due to the higher nuclear thresholds~\cite{Haxton:1987kc, Fukugita:1988hg, Engel:1996zt, Kolbe:1999au, Hayes:1999ew, Volpe:2000zn, Auerbach:2001hz, Kolbe:2002gk, Laha:2013hva, Laha:2014yua, Lu:2016ipr}.

DUNE is a next-generation long-baseline neutrino experiment~\cite{Abi:2020evt}, currently in the design phase.  Its far detector will ultimately have a fiducial mass of 40~kton (in four detector modules) and will detect charged particles using the liquid argon time-projection chamber technique.  In our calculations, we assume the full 40-kton mass, even though DUNE will initially start with just two modules.  While DUNE is primarily intended for GeV-range neutrinos, it will also have important capabilities at lower energies.  For MeV-range neutrinos, there are many new, challenging aspects about detection---cross section, detector response, backgrounds, and more---with associated uncertainties~\cite{Acciarri:2015uup, Capozzi:2018dat, Zhu:2018rwc, Abi:2020evt}.

In our calculations, we consider only the $\nu_e + \, ^{40}\mathrm{Ar} \rightarrow e^- + \, ^{40}\mathrm{K}^*$ interaction, which isolates the $\nu_e$ flux.  Even though the yields for neutrino-electron scattering are dominated by $\nu_e$, other flavors contribute, so these events must be separated.  (Neutrino-electron scattering events are much more important in DUNE than in Super-K because the neutrino-argon cross section is small and the inverse-beta cross section is large.)  The simplest method to isolate the neutrino-electron scattering events is to take advantage of their very forward kinematics, which is in contrast to the near-isotropic neutrino-argon scattering events.  Removing events within a forward cone of half-angle $40^\circ$ leaves $\simeq$90\% of the neutrino-argon events while removing nearly all of the neutrino-electron scattering events~\cite{Arneodo:2000fa}.  Given the large detector-related uncertainties for DUNE, we neglect the modest penalty factor.

\begin{figure*}[t]
\centering
\includegraphics[width=\columnwidth]{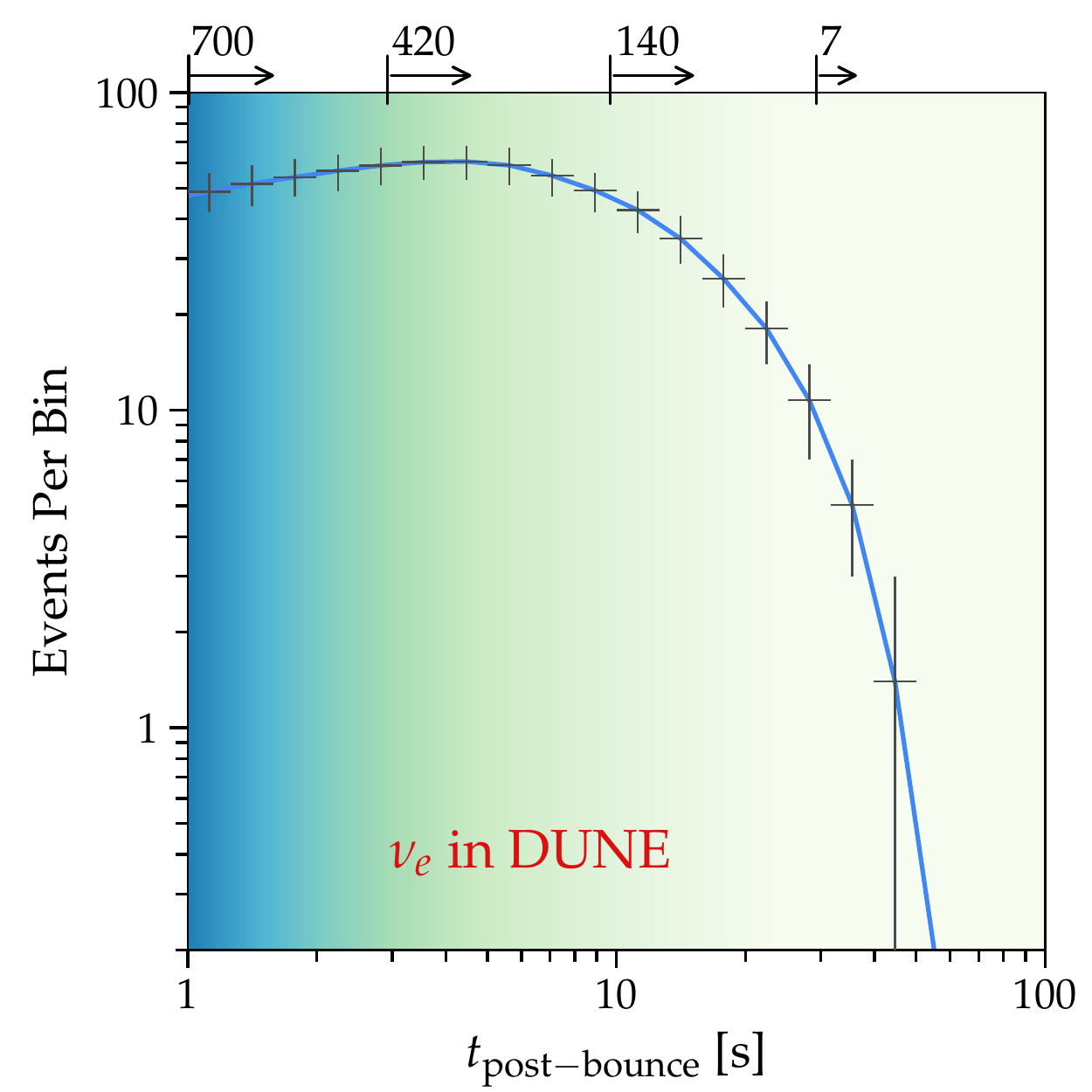}
\includegraphics[width=\columnwidth]{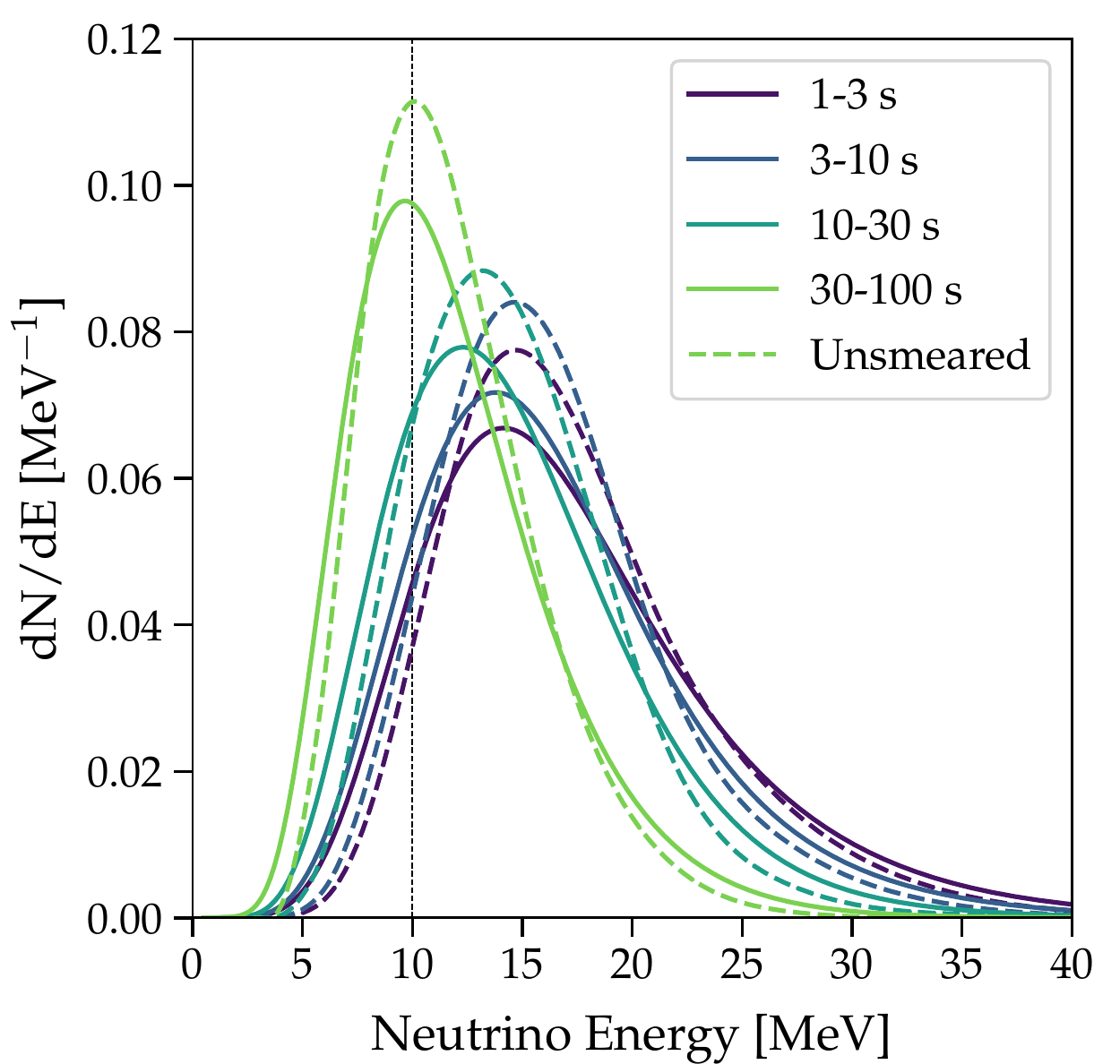}
\caption{Similar to Fig.~\ref{fig:spectrum_DUNE} for DUNE, but for the second case, based on neutrino energy.
{\bf Left:} Event rate for a threshold of 10~MeV in neutrino energy.
{\bf Right:} Solid and dashed lines show the smeared and true neutrino energy spectra. The vertical dotted line at 10 MeV indicates the assumed DUNE threshold.}
\label{fig:spectrum_DUNE_nu}
\end{figure*}

The cross section for $\nu_e + \, ^{40}\mathrm{Ar} \rightarrow e^- + \,^{40}\mathrm{K}^*$ is not measured experimentally, nor is it well known theoretically.  The key input data on the transition strengths come from $(p,n)$ and beta-decay measurements~\cite{Bhattacharya:2009zz}.  For reviews, see Refs.~\cite{Capozzi:2018dat, Gardiner:2018zfg}.  The Q-value for the interaction to the ground state of $^{40}\mathrm{K}$ is 1.504 MeV, but this transition is forbidden by selection rules.  The cross section thus consists of transitions to a variety of excited states, of which the most important is $\Delta E_i = 4.384$~MeV above the ground state.  These excited states decay through gamma-ray emission, leading to secondary-electron production through Compton scattering and pair creation.  We calculate the cross section taking into account the 15 most important excitations.  For each state $i$, the primary electron kinetic energy is $T_e = E_\nu - Q_i$, where $Q_i = 1.504~\mathrm{MeV} + \Delta E_i$.  (For the most important state, $T_e = E_\nu - 5.888$~MeV.)  As discussed in Ref.~\cite{Capozzi:2018dat}, this treatment works best for neutrino energies below about 15--20 MeV.  At higher neutrino energies, the cross section is more complex and uncertain, but including more transitions would only increase the cross section relative to our calculation.  These transitions often involve nucleon emission, which we neglect.  According to {\tt MARLEY}~\cite{Gardiner:2018zfg, MARLEYv1.2.0}, for the spectrum between 1--10~s, there is an 8\% increase to the cross section if we add neutron-emitting transitions, and between 10--60~s, there is a 5\% increase.

A key question is the fraction of the neutrino energy in an event that DUNE can observe, which depends on the detectability of the secondary electrons produced by nuclear gamma rays.  We present results for two cases and then discuss the prospects in detail.  In the first case, we assume that only the primary electron is measured, with its kinetic energy $T_e$ smeared by the energy resolution, and that all de-excitation gammas, corresponding to a total energy $\Delta E_i$, are lost. For realistic event reconstruction, this would be similar to the scenario where the primary electron charge track is detected in the time-projection chamber, but the isolated short tracks from secondary electrons (``blips") are not measured.  (The distinction is that we ignore bremsstrahlung of primary electrons, which becomes important at higher energies.)  This assumption is inspired by the DUNE Conceptual Design Report~\cite{Acciarri:2015uup} and Refs.~\cite{Capozzi:2018dat, Zhu:2018rwc}.  The abundance of nuclear excited states makes the mapping between neutrino energy and the primary electron energy multivalued, so the neutrino spectrum must be fit through forward modeling.  We assume a 20\% energy resolution for the primary electrons and that events can be detected with perfect efficiency above 6 MeV electron kinetic energy (and zero at lower energies).  We first show results for the first case, then detail the second case.

Figure~\ref{fig:spectrum_DUNE} (left panel) shows the event rate of the $\nu_e$ neutrino-argon interaction in DUNE for the first case (using primary electron energy), assuming an electron-energy threshold of 6 MeV.  The shape of the curve is again similar to that of the luminosity.  (An important signature for the $\nu_e$ channel is the neutronization burst~\cite{Kachelriess:2004ds,Chiu:2013dya,deGouvea:2019goq,Abi:2020lpk}, which we do not discuss because it occurs at early time, $t\simeq 0$~s.) The drop in the event rate around 40 s is due to both the decreasing number flux and average energy (and hence cross section), but the change in the number flux dominates.  The overall event yield in DUNE is lower than in Super-K, but is still high, with 110 events expected in DUNE after 10 s.  The last event would be detected at $\simeq$40 s, in principle late enough to probe much of the transition of the PNS to neutrino transparency.  The gray dashed line shows the most significant background in the MeV range, which is induced by thermal captures of external neutrons from radioactivities in the rock surrounding DUNE~\cite{Arneodo:2000fa, Capozzi:2018dat, Zhu:2018rwc}.  These neutrons uniformly fill the volume, and their captures produce gamma rays and thus electrons.  We calculate this background rate by simulating the injection of neutron-capture gammas into liquid argon and recording the energy of individual electrons, then smearing the electron spectrum with 20\% energy resolution~\cite{Capozzi:2018dat, Zhu:2018rwc}.  The background rate is about 4 Hz, crossing the signal rate at $\simeq$20~s.  This background could be dramatically reduced with even a modest amount of shielding~\cite{Capozzi:2018dat, Zhu:2018rwc}.  Varying the analysis threshold drastically changes the background rate. With a threshold of 5 MeV, the crossing would instead be at $\simeq$10~s; for 7 MeV, it would be at $\simeq$30~s.  (The 20\% energy resolution is conservative; with 7\% resolution and thresholds of 5, 6, and 7 MeV, the crossings would be at $\simeq$20 s, 40 s, and late enough to be irrelevant.)
 
Figure~\ref{fig:spectrum_DUNE} (right panel) shows the time-binned electron energy spectra in DUNE.  Compared to the $\bar{\nu}_e$-induced spectra in Super-K, here the prospects are less good, because the counts are lower and because the spectrum peaks closer to the detector threshold.  The energy resolution is taken into account but has negligible effects on the appearance of the spectrum. The reasons that the $\nu_e$ spectrum is more challenging to measure are that the average energy is lower, the neutrino interaction and detector thresholds are higher, and the relationship between neutrino and electron energy is less favorable.  With only 5 expected events between 30--100~s and the spectrum peak below threshold, only minimal information will be available.  Still, some tests of the spectra should be possible at earlier times.  Improving the threshold by lowering backgrounds is of critical importance to the spectrum because any spectrum fit degrades dramatically when the peak is near threshold.  In contrast, lowering the threshold is less important for the time profile.

In the second case, we assume that the full neutrino energy can be reconstructed with no shift, but subject to energy resolution.  This case, which is more optimistic, is inspired by progress in detecting blips in small liquid-argon detectors~\cite{Acciarri:2018myr, Foreman:2019dzm, Castiglioni:2020tsu} and by the projections for photon detection made in the recent DUNE Technical Design Report~\cite{Abi:2020evt}.  In either case, the full neutrino energy would be estimated from the sum of the primary electron energy, the combined energy of the secondary electrons, and 1.504 MeV for the ground-state $Q$ value.  Except for measurement uncertainties and neutron emission, the energy mapping is single-valued because the nuclear transition of each interaction would be identified.  We assume 20\% energy resolution for the neutrino energy and that events can be detected with perfect efficiency above 10 MeV neutrino energy (and zero at lower energies).  Detecting secondary electrons might enable clean separation between neutrino-argon and neutrino-electron events~\cite{Castiglioni:2020tsu}.

Figure~\ref{fig:spectrum_DUNE_nu} shows the rate and spectrum for $\nu_e$ events in DUNE for the second case (using neutrino energy). We adopt a threshold of 10~MeV for the event rate calculation, which is roughly comparable to a threshold of 4~MeV on electron energy, making the event rate similar to that in Fig.~\ref{fig:spectrum_DUNE}.  The spectrum is improved by shifting the signal to higher energies while leaving the background unchanged.  (The 20\% smearing effects are now visible due to the larger absolute energies compared to the first case.)  However, the quality of the spectrum reconstruction between 10--30~s is still likely not very good because the spectrum peak is close to threshold.  If the threshold is reduced to below 7 MeV, the neutron background becomes overwhelming at late times.  (The neutron-capture gamma rate is 5 Hz above 7 MeV and 0.6 Hz above 8~MeV.) Ideally, it will be possible to reject the neutron background by shielding or through the event topology, both of which should be seriously explored~\cite{Capozzi:2018dat, Zhu:2018rwc}.

DUNE's MeV capabilities remain uncertain.  Decisive steps are needed to maximize these, as outlined in Refs.~\cite{Capozzi:2018dat, Zhu:2018rwc}.  The neutrino-argon cross section must be measured, which is feasible.  The critical neutron-capture background must be reduced through shielding or perhaps particle identification, which seems quite promising~\cite{Castiglioni:2020tsu}.  Other backgrounds must be reduced through material selection and purification.  Realistic event reconstruction studies need to be carried out.  The energy resolution must be improved through better charge reconstruction or more extensive instrumentation to collect scintillation light.  These points are general to DUNE's MeV physics prospects, including solar neutrinos, supernova neutrinos, and event reconstruction for GeV-range interactions.  With such potential broad scientific impacts, DUNE's MeV efforts deserve significantly more resources for personnel and hardware.

In summary for the DUNE measurement of $\nu_e$, the signal count rate is high and spectra could be measured, especially if the gamma rays from nuclear de-excitation can be detected.  If DUNE were as successful as it should be, then its $\nu_e$ results would be critical for testing PNS cooling to late times.  In combination with Super-K's $\bar{\nu}_e$ results, one could test if the luminosities and average energies of these two flavors have converged, a key prediction. This is also essential for measuring the total deleptonization of a SN.  However, even this is not enough, as we discuss next.


\subsection{$\nu_x$ in JUNO}
\label{subsec:JUNO}

\begin{figure*}
\centering
\includegraphics[width=\columnwidth]{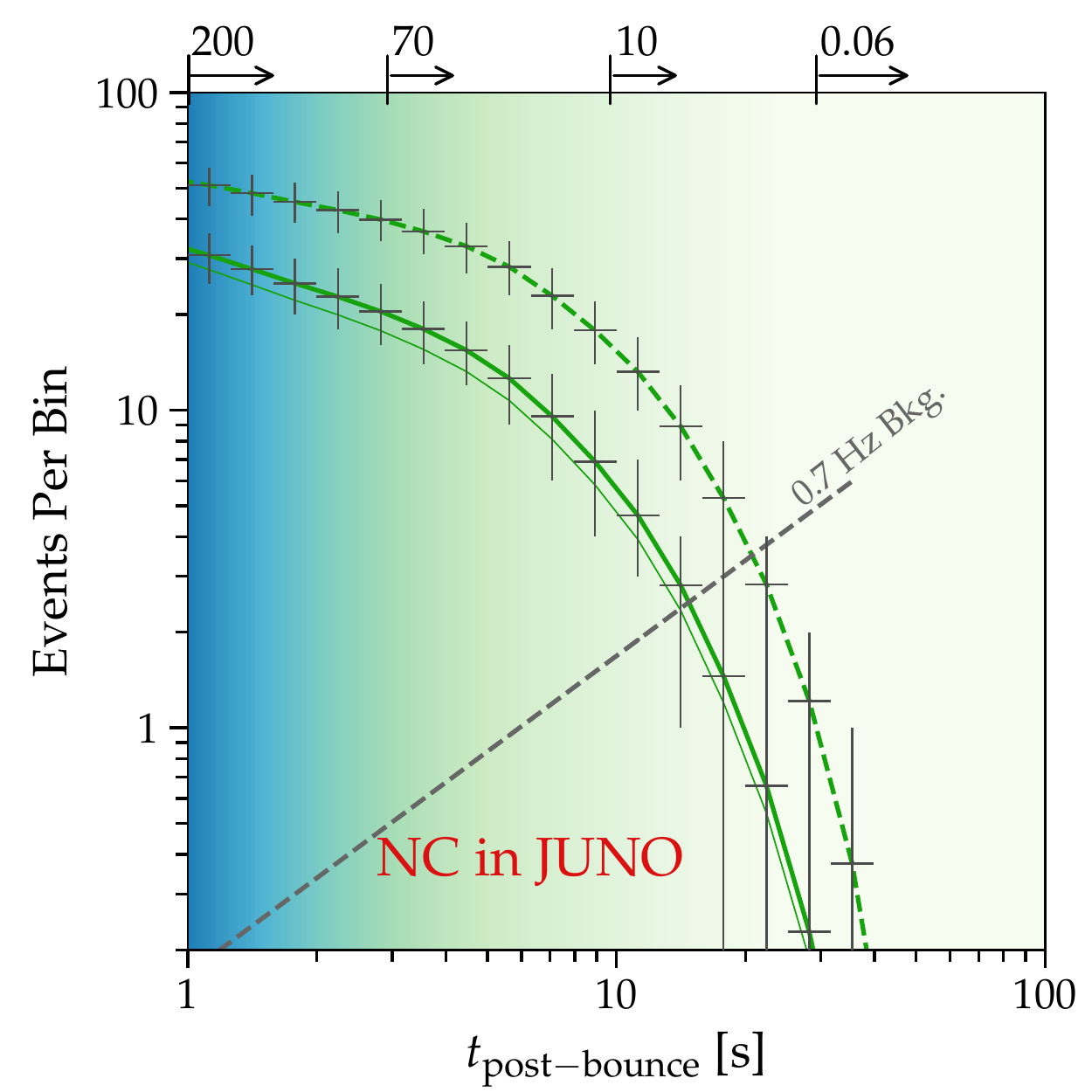}
\includegraphics[width=\columnwidth]{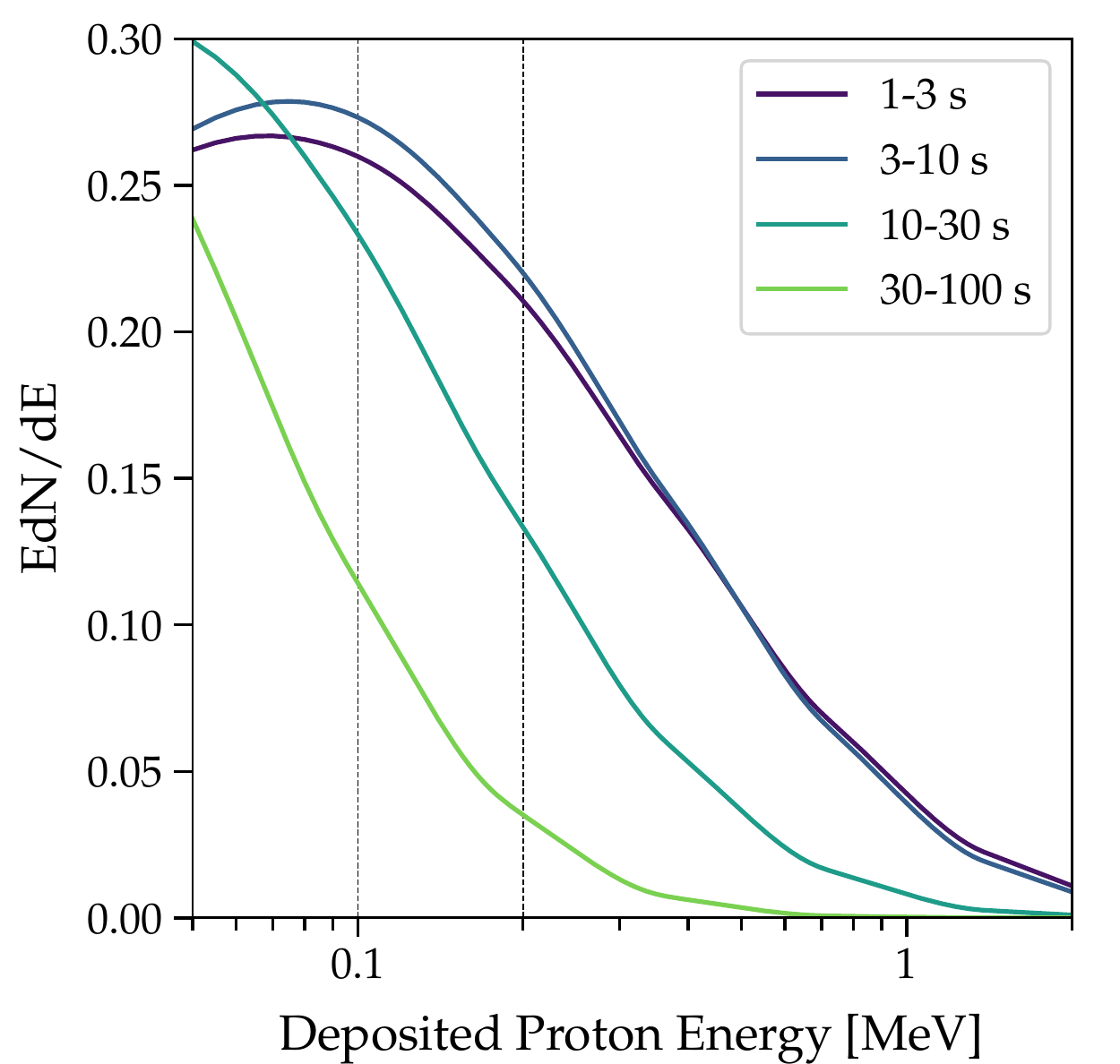}
\caption{Similar to Fig.~\ref{fig:spectrum_SK}, but for the neutral-current events in JUNO due to neutrino-proton elastic scattering.
{\bf Left:} Event rate for two cases: a solid line for a detection threshold of 0.2 MeV (nominal design) and a dashed line for a detection threshold of 0.1 MeV (optimistic).  The counts on the top axis and the background rate are both for the 0.2-MeV case. The thin solid line shows the $\nu_x$ contribution for the 0.2~MeV case.
{\bf Right:} For the spectra, we plot $E \, dN/dE = dN/d\ln{E} = (2.3)^{-1} \, dN/d\log{E}$ to match our log-$E$ axis.  The vertical dotted lines at 0.2 MeV and 0.1 MeV indicate the possible thresholds.}
\label{fig:spectrum_JUNO}
\end{figure*}

For $\nu_x$, the best detector will be JUNO~\cite{An:2015jdp}, using elastic scattering on free (hydrogen) protons, 
\begin{equation}
\nu + p \rightarrow \nu + p, 
\end{equation}
which, as a neutral-current interaction, proceeds for all six flavors~\cite{Beacom:2002hs, Dasgupta:2011wg}.  Only low-threshold scintillator detectors like JUNO can detect the proton recoil.  In Super-K, DUNE, and JUNO, there are neutral-current interactions with nuclei~\cite{Raghavan:1986fg, Fukugita:1988hg, Langanke:1995he, Engel:1996zt, Armbruster:1998gk, Kolbe:2002gk}, but the yields are small, especially with the low expected average neutrino energies at late times, so we neglect them. (See discussions of $\bar\nu_e$ detection in JUNO in Sec.~\ref{subsec:SK}.)

JUNO is a next-generation reactor neutrino experiment, starting in 2022~\cite{JUNONu2020}, designed to make precision measurements of neutrino mixing~\cite{An:2015jdp}.  Its fiducial mass will be 20 kton of liquid scintillator, viewed by photomultiplier tubes mounted on the surface.  JUNO will measure the kinetic energy depositions of charged particles through the scintillation light they produce.  Due to the high light yield in JUNO, the energy resolution is superb, $\delta E/E \simeq 3\%/ \sqrt{E(\mathrm{MeV})}$.  The nominal energy threshold is 0.2 MeV, but we also calculate results for an optimistic threshold of 0.1 MeV.  We assume perfect efficiency above threshold and zero below.

For nonrelativistic particles like the recoil protons from the $\nu + p \rightarrow \nu + p$ interaction, their high energy loss rate leads to a reduction of their light output (``quenching") relative to an electron of the same energy.  The electron-equivalent energy $T'$ can be calculated by
\begin{equation}
T'(T) = \int_0^T \frac{dT}{1+k_B\langle dT/dx\rangle} ,
\end{equation}
where $k_B$ is Birks' constant, for which we take $k_B = 0.0098$~cm/MeV~\cite{An:2015jdp, vonKrosigk:2013sa}.  For example, protons of true energies 2.0, 1.0, and 0.5 MeV register with energies smaller by factors of 3.5, 5, and 6.5.

The cross section for $\nu + p \rightarrow \nu + p$ is large and is well known theoretically~\cite{Beacom:2002hs, Dasgupta:2011wg}.  There is no threshold.  The proton has a distribution of energies from zero up to a kinematic maximum $\simeq$ $2 E_\nu^2 / M_p$ (in our calculations, we use the exact expression, as Ref.~\cite{Bar:2018ejc} showed that the steep spectrum makes approximation inaccurate).  The differential cross section favors the largest allowed recoil energies because the cross section is dominated by the axial-vector response due to the near-cancellation of the vector response.  The cross section is independent of flavor, except that it is somewhat different for neutrinos and antineutrinos due to weak-magnetism corrections.  However, those differences are small at the low neutrino energies we consider and largely cancel when summing over all flavors.  We neglect possible cross-section contributions due to strange quarks in the proton.

Figure~\ref{fig:spectrum_JUNO} (left panel) shows the total event rate of neutral-current neutrino-proton scattering in JUNO.  It is strongly dominated by the $\nu_x$ contribution, shown as the thin green line for the 0.2 MeV threshold case, complementary to the previous two channels.  The shape of the curve drops more steeply than that of the luminosity.  The principal reason is the effect of the detector threshold.  The total yield from neutrino-proton elastic scattering is actually comparable to that from the inverse-beta interaction, because the somewhat smaller total cross section is compensated by the larger number of participating neutrino flavors.  But, due to quenching, the detectable proton recoil spectrum is very steeply falling, and most of the events are below the detector threshold.  As the neutrino average energies fall, the maximum proton energies decrease quadratically and the quenched proton energies decrease even faster, so the number of events above threshold falls steeply.  For a 0.2~MeV threshold, we expect only 10 events after 10~s, with the last event at $\simeq$20~s.  For a 0.1~MeV threshold, we expect only 30 events after 10~s, with the last event at $\simeq$30~s.

Neutrino-proton elastic scattering events will be easily separated from other core-collapse neutrino interactions, which produce higher-energy depositions.  Further, inverse-beta events will be isolated by the coincident detection of their positrons and neutrons, charged-current interactions with $^{12}\mathrm{C}$ by their  subsequent nuclear decays, and neutral-current interactions with $^{12}\mathrm{C}$ by their 15.11-MeV gamma ray.  The only possible concern is neutrino-electron scattering events.  However, the yields are not large, the fraction of events at the lowest energies is small, and pulse-shape discrimination should help.

Detector backgrounds are a limiting factor to a lower threshold. The 0.7 Hz background rate above 0.2~MeV shown in Fig.~\ref{fig:spectrum_JUNO} is from $^{85}$Kr beta decay, assuming an activity of 50~$\mu$Bq/m$^3$~\cite{An:2015jdp}. This background rate does not increase much for a lower threshold.  A more serious concern is $^{14}$C beta decay, which has an endpoint of 0.156~MeV. Assuming a concentration of $10^{-17}$g/g, there is negligible background above 0.2~MeV, 1.4~Hz above 0.17~MeV, and 50~Hz above 0.15~MeV. Lowering $^{14}$C contamination through material selection is critical to pushing the detection threshold toward 0.1~MeV~\cite{Fang:2019lej}.  Pulse-shape discrimination techniques may also help.

Figure~\ref{fig:spectrum_JUNO} (right panel) shows the time-binned proton detected energy spectra in JUNO.   Energy resolution is included, but has little effect.  We see the critical role of the energy threshold for neutral-current detection.  With a threshold of 0.2 MeV, we cannot recognize the peak of the spectrum.  After a few seconds, we sample only far into the tail.  With a threshold of 0.1 MeV, the situation would be somewhat better.

In summary for the JUNO measurement of $\nu_x$, the signal count rate is low but it may be adequate, especially if the threshold can be further lowered.  The most important point is that the count rate and spectrum should be well enough measured to at least a few seconds to test if the luminosities and average energies of all flavors are converged.  Once they converge, it is likely that they stay converged unless something dramatic happens, which could be visible in the high-statistics $\bar{\nu}_e$ count rate.   Neutral-current measurements are also crucial to test for neutrino mixing (i.e., before the neutrino spectra converge) and to detect possible late-time accretion, which we discuss below. 


\section{Physics Prospects}
\label{sec:physicsresults}

In this section, we first discuss neutrino mixing, then highlight examples of the physics opportunities that arise from detecting late-time core-collapse neutrinos.  We consider tests of core-collapse physics, how to distinguish NS and BH outcomes, and how to measure the transition time at which the PNS becomes a NS or BH.  Further work is needed to realize the full potential of this data.


\subsection{Testing for neutrino-mixing effects}

Neutrino mixing inside supernovae is an unsolved theoretical problem, and it affects how one would extract physics results from data. In this subsection, we briefly outline how one may test the presence or the size of mixing effects from data while being agnostic about the details of the mixing mechanisms.  We emphasize that quantitative, realistic analyses are needed to investigate all the possibilities and to assess the impact of mixing on a variety of physics questions.

We first note that quantities that depend on sums of flavors, e.g., the total neutrino luminosity, are less susceptible to uncertainties in mixing effects than quantities that depend on flavor differences, e.g., the total deleptonization rate. In addition, if the initial fluxes and spectra of two flavors are similar, then mixing between them can have only small effects.  This is expected to be the case for all flavors during the late cooling phase and for $\bar{\nu}_e$ and $\nu_x$ during the early cooling phase.

For probing neutrino spectra, neutral-current channels have a special role because they are equally sensitive to all neutrino flavors (neglecting some small differences between neutrinos and antineutrinos) and are thus insensitive to active-flavor mixing.  The $\nu + p \rightarrow \nu + p$ channel in JUNO is especially important because this is the only neutral-current channel with differential spectrum information for $\nu_x$~\cite{Beacom:2002hs, Dasgupta:2011wg}.  Beyond the mixing independence, there is a deeper point about this channel~\cite{Laha:2013hva, Laha:2014yua}. If the initial neutrino spectra are different, with the $\nu_x$ spectrum significantly hotter than the others, then the $\nu + p \rightarrow \nu + p$ yields are strongly dominated by $\nu_x$, because the total cross section, differential cross section, and quenching effects all favor the highest-energy neutrinos~\cite{Beacom:2002hs, Dasgupta:2011wg}.  An example is shown above in Fig.~\ref{fig:spectrum_JUNO}.  In this case, the $\nu + p \rightarrow \nu + p$ measurements would isolate the {\it initial} $\nu_x$ spectrum, regardless of its labels at the detector.  This is a key insight.

To test for the presence of mixing, we can compare measurements of the effective spectra at Earth using the $\bar{\nu}_e + p \rightarrow e^+ + n$ channel in Super-K, the $\nu_e + \, ^{40}\mathrm{Ar} \rightarrow e^- + \, ^{40}\mathrm{K}^*$ channel in DUNE, and the all-flavor $\nu + p \rightarrow \nu + p$ channel in JUNO.  We illustrate the qualitative possibilities by considering some extreme scenarios for the early PNS cooling phase, during which the $\nu_e$ average energy is expected to be $\simeq$2--6~MeV below that of $\nu_x$ and $\bar{\nu}_e$ for most simulation results, e.g., as in Refs.~\cite{Hudepohl:10, Fischer:11, Roberts:12}.  (This can be generalized to the explosion phase, during which all three flavors are expected to have different average energies.)  If the $\nu_e$ average energy measured from the effective spectrum is indeed lower than that of the other flavors, then mixing effects are likely small and the neutral-current signal is dominated by five flavors.  If, instead, the measured $\nu_e$ average energy is comparable to that of the other flavors, there are two possibilities.  It could be that all initial spectra were similar (i.e., converged), in which case the neutral-current signal would reflect the contributions of six flavors.  Or it could be that there were strong mixing effects, in which case the neutral-current signal would reflect the contributions of just five flavors.  Beyond the conceptual tests outlined here, one can also look for spectrum features.

How likely we are to succeed at disentangling mixing from the intrinsic properties of the neutrino emission depends on how well the detectors perform.  That is why it is so critical that experiments work to achieve lower detection thresholds, better background rejection, and more accurate cross section predictions.  New techniques to better extract spectra, e.g., as in Ref.~\cite{Nagakura:2020bbw}, should be developed further.  In addition, continued theoretical work on both mixing and PNS evolution is needed.


\subsection{Probing the physics of core collapse}

Neutrino observations of a Milky Way event have the potential to decisively answer many longstanding questions about the complex physics of core collapse.  Compared to SN 1987A, with 19 events, we expect huge yields, $\simeq$$10^4$ events ($\simeq$$10^5$ once we have Hyper-K).  But yields are not the whole story---success depends upon having near-complete coverage in flavor, energy, and time, and we may only have one chance to get this right.  Here we discuss some of the most fundamental questions that may be answered with such a detection, assuming that neutrino mixing will be well enough understood.

Observations during the long PNS cooling phase, which begins within the first seconds after core bounce, are especially important for tests of the total emitted energy ($\propto M_{\rm NS}^2 / R_{\rm NS}$) and lepton number ($\propto M_{\rm NS}$), assuming we know the distance (Appendix~\ref{sec:distance}).  These integral quantities directly reflect the net effects of core collapse, probing the NS mass and radius~\cite{Prakash:97, Pons:99}.  Many details of the PNS physics (e.g., neutrino opacities), while greatly changing the {\it rates} of neutrino emission, may change these integral quantities much less.  Quantitative exploration of this conjecture is needed.

The fractions of energy and lepton number emitted during the PNS cooling phase are significant.  Figure~\ref{fig:nominal_number} shows that in this simulation, $\simeq$50\% of the energy and $\simeq$30\% of the lepton number are emitted after 2 s, which we nominally take to be the start of PNS cooling.  Even after 10 s, the fractions are $\simeq$20\% and $\simeq$10\%, respectively.  As shown in Sec.~\ref{sec:resultsNS}, it should be possible to measure $\bar{\nu}_e$ events to $\simeq$50 s, $\nu_e$ to $\simeq$20 s, and $\nu_x$ to $\simeq$20~s under conservative assumptions about detector backgrounds, and ideally longer, though with only low statistics at the latest times.

Measurements of the differential event rates are important for probing the details of PNS models, not just for constructing the integral quantities that characterize the properties of the final NS remnant.  At earlier times, testing differences between the $\nu_e$ and $\bar{\nu}_e$ event rates and spectra probes the deleptonization process, the timescale of which is sensitive to the nuclear equation of state and neutrino opacities~\cite{Roberts:12, Roberts:12a, Martinez-Pinedo:12, Nakazato:2012qf, Suwa:2019svl, Nakazato:2019ojk, Nakazato:2020ogl, Fischer:20}, as well as nucleosynthesis in the neutrino-driven wind, which forms the innermost portion of the supernova ejecta and is sensitive to the difference between the $\nu_e$ and $\bar{\nu}_e$ average energies~\cite{Arcones:11}.  In addition, an excess of $\nu_e$ and $\bar{\nu}_e$ emission relative to $\nu_x$ would indicate substantial accretion, as predicted at late times in some models with rather different luminosity evolution~\cite{Blum:2016afe, Bar:2018ejc}.

At later times, measurement of the cooling timescale probes the properties of warm, dense, neutron-rich matter. This includes the nuclear equation of state~\cite{Pons:99, Roberts:12, Suwa:2019svl, Nakazato:2019ojk, Nakazato:2020ogl}, neutrino opacities~\cite{Reddy:99, Hudepohl:10}, convection in the PNS interior (sensitive to the equation of state)~\cite{Roberts:12}, the possible presence nuclear pasta-like structures in the outer layers of the PNS at late times~\cite{Horowitz:16, Roggero:18}, and possible beyond-the-standard-model cooling processes~\cite{Raffelt:96}.  For more details, see Sec.~\ref{subsec:variation}.  Although it may be challenging to disentangle the impact of various processes inside the PNS that affect the cooling timescale, measurement of the cooling timescale is likely to be relatively insensitive to neutrino mixing. Complementary to our focus on the overview and the detection aspects, Refs.~\cite{Suwa:2019svl, Nakazato:2019ojk, Nakazato:2020ogl, Suwa:2020nee} have carried out extensive studies of PNS cooling signals with different inputs.

How well these questions can be answered strongly depends on the experimental performance of Super-K, DUNE, and JUNO.  For the questions just discussed, all three are needed to ensure complete flavor coverage.  Next, we explore two questions that can be answered even if only one flavor can be measured well.  As an example, we focus on $\bar{\nu}_e$ detection in Super-K.


\subsection{Distinguishing neutron-star and black-hole outcomes}

A key question is the outcome of the explosion: does the PNS eventually become a NS or a BH?  At early times, the luminosity (and the event rate) falls approximately as $1/t$.  For the NS case, a distinctive signature is the steep drop in flux relative to this trend, which signals the onset of neutrino transparency.  For the BH case, the defining signature is a sharp truncation in the flux.  With low statistics, these cases can be hard to distinguish; e.g., we still do not know the final outcome of SN 1987A.  With Super-K, which has a much greater mass and much lower backgrounds compared to the SN 1987A detectors, we expect robust detections until very late times.

We separately consider the different scenarios of Fig.~\ref{fig:BH_formation_luminosity}.  The clearest case is Case 2, where a BH forms early, so that the event rates are large until they are sharply truncated to zero, which cannot be confused with other outcomes.  The more challenging question is whether we can distinguish Case 1, the formation of a NS, from Case 3, the formation of a BH at late times.  A quantitative statement depends on what we assume for the luminosity before the transition.  For the NS outcome, we use our nominal PNS model.  For the BH outcome, to be conservative, we assume that truncation of the flux occurs relative to the flux of this PNS mode, instead of assuming that the $1/t$ trend continues.

$\bullet$
To identify NS formation, one has to see the expected steep drop in the neutrino luminosity relative to the $1/t$ trend.  If detector backgrounds are vanishing, as expected for Super-K, this is easy.  Then, to distinguish this from the BH case, one only needs to confirm that the luminosity does not decrease to zero. At a given time, this can be confirmed with infinite significance as long as there are still detected signal events, because the BH case would have zero.  The BH-formation case can thus be rejected out to the time of the last expected event, $\simeq$50~s in our nominal PNS cooling model.  After this, the NS could collapse to a BH without our knowing.

\begin{figure}[t]
\centering
\includegraphics[width=1.05\columnwidth]{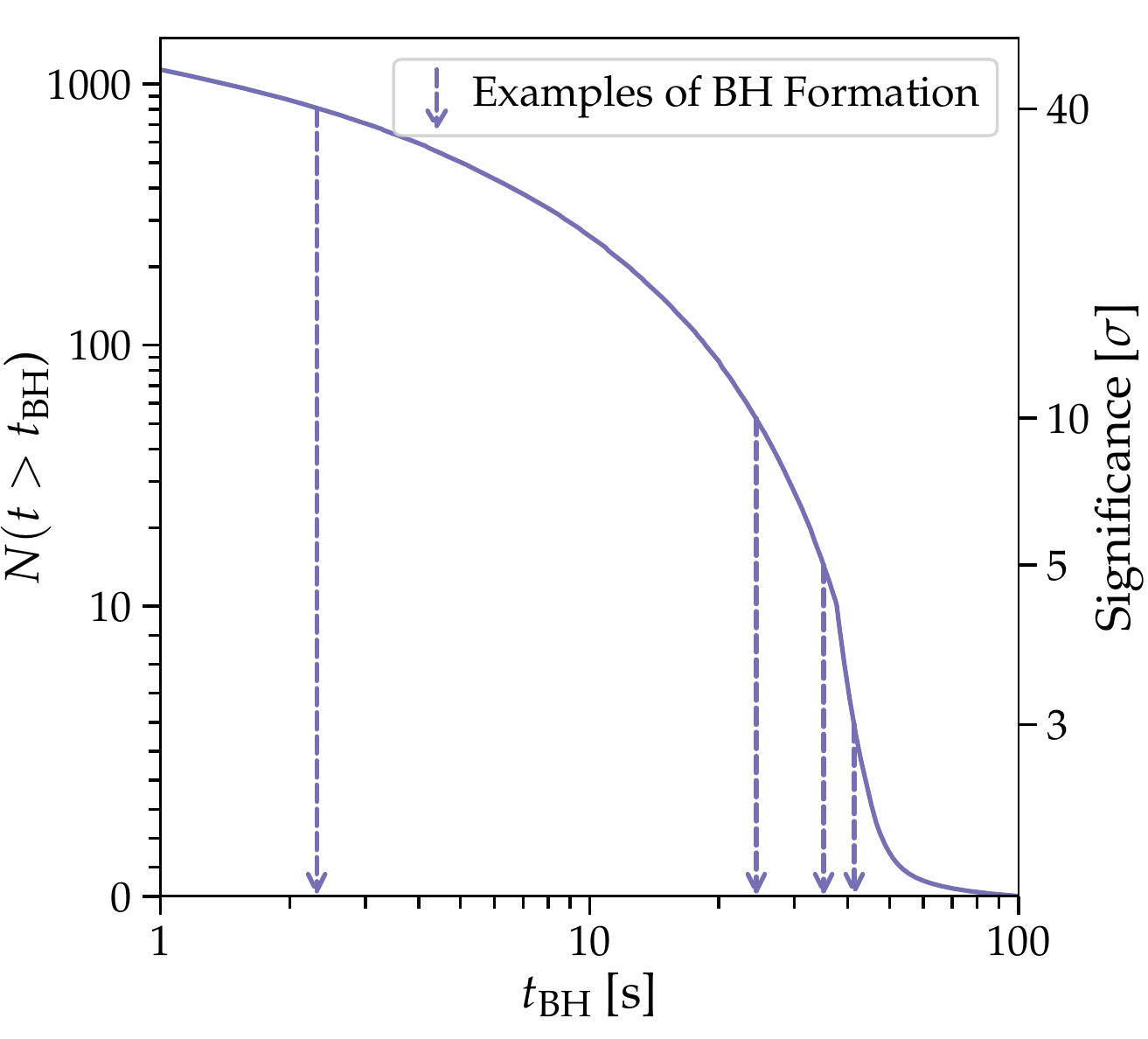}
\caption{Prospects for detecting BH formation in Super-K, showing the parameter dependence on $t_{\rm BH}$.  The left $y$-axis shows the number of events after the BH formation time, $t_{\rm BH}$, {\it that would have been expected in our nominal PNS model}.  The right $y$-axis shows the corresponding significance level for identifying BH formation. The vertical lines mark the times corresponding to detection significances of 40, 10, 5, and 3$\sigma$.}
\label{fig:BH_detection}
\end{figure}

$\bullet$
To identify BH formation, one has to see a sharp truncation of the neutrino luminosity.  At the latest times, when the expected event number is low, this could be mimicked by the NS case with a downward Poisson fluctuation in the event rate.  We use our nominal model as the default flux, and vary the BH formation time $t_{\rm BH}$ by setting the flux to zero after that.  Figure~\ref{fig:BH_detection} illustrates the results. We compute how many events we would have expected after $t_{\rm BH}$ if there had been no BH formation (shown in the left $y$-axis in Fig.~\ref{fig:BH_detection}), and how likely it is for this expected number to fluctuate down to zero, which we define as the significance of a BH detection (shown in the right $y$-axis).  As expected, if the signal truncates when the luminosity is high, BH formation can be confirmed with very high significance. Even if the signal truncates at 40~s, where the luminosity is already low, BH formation can still be identified at 3$\sigma$.

Thus, for a core collapse at 10 kpc, Super-K should be able to measure the formation of the remnant and identify it as a NS or BH with high significance.


\subsection{Measuring the transition time}

Once the formation and nature of the remnant are confirmed, the next question is how well the transition time can be measured.  This is important for distinguishing models, as it depends on the progenitor structure, NS equation of state, and more.

\begin{figure}[t]
\centering
\includegraphics[width=\columnwidth]{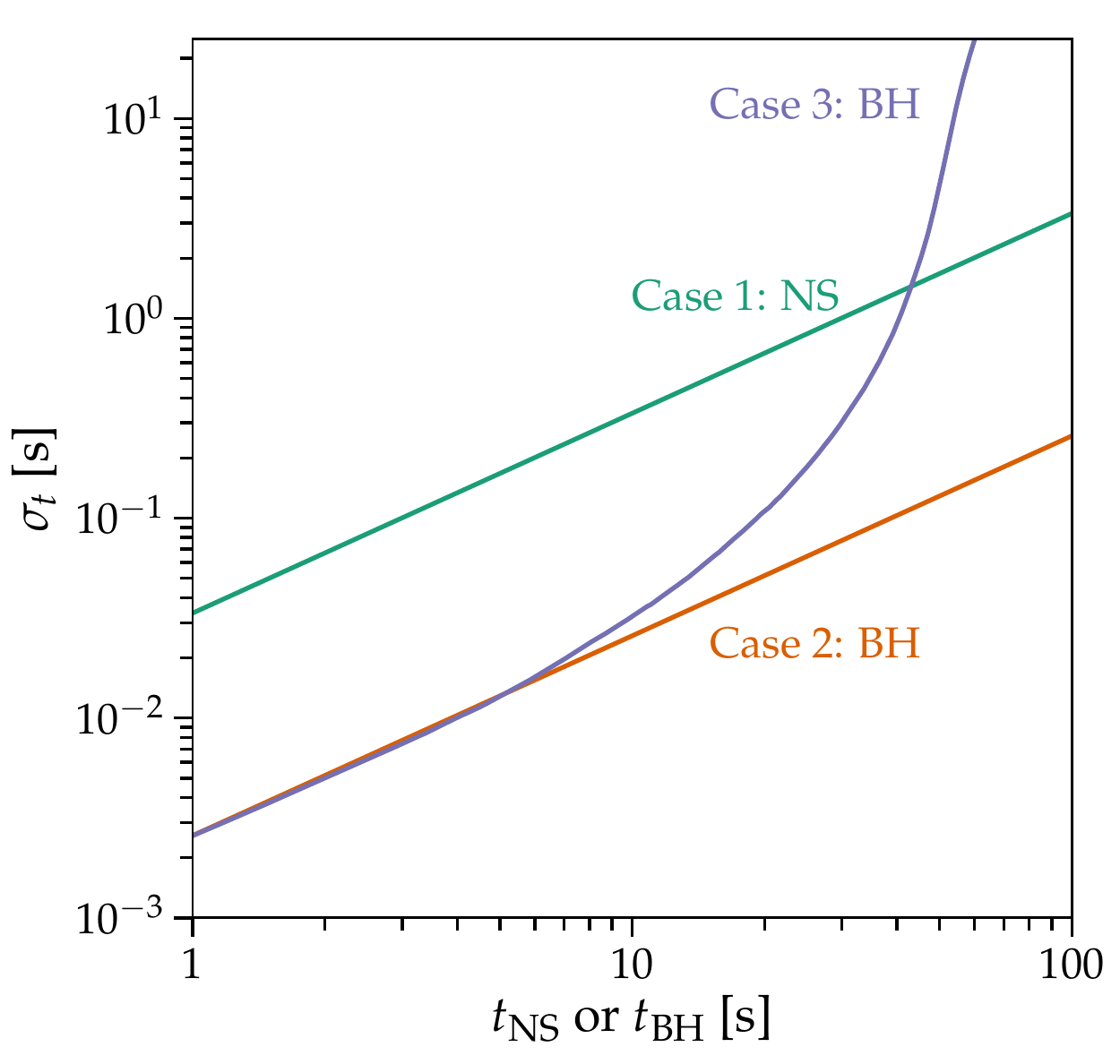}
\caption{Expected measurement precision for the transition time.  For Case 1 (NS formation), this is the transparency time, $t_{\rm NS}$, when there is a super-exponential drop relative to the $1/t$ trend.  For Case 2 (early BH formation), this is the BH formation time, $t_{\rm BH}$, where we assume the sharp truncation occurs relative to the $1/t$ trend.  For Case 3 (late BH formation), we assume the sharp truncation occurs relative to the super-exponential drop of the NS-forming case.}
\label{fig:critical_time}
\end{figure}

Similar to distinguishing NS and BH formation, the precision of the measured transition time depends on what we assume before the transition.  Ideally, one should take into account the time variation of both the luminosity and average energy, compute the detected event rate and spectra, infer the average energy from detected spectra, and then fit for the luminosity change.  Because we are aiming for an approximate understanding of the expected precision and the average energy evolution is moderate in our nominal model, we instead consider simpler scenarios that resemble those in Fig.~\ref{fig:BH_formation_luminosity}.

Figure~\ref{fig:critical_time} shows the expected precision of the measured transition time, i.e., transparency time for a NS or formation time for a BH, as a function of its true value.

$\bullet$
For NS formation (Case 1), we assume that the event rate follows $(1/t) \times e^{-(t/t_{\rm NS})^2}$, where $t_{\rm NS}$ is 36 s in our nominal model and it occurs in tens of second for most model scenarios, and where we fit $t_{\rm NS}$ directly from the event rate.  The precision $\sigma$ has a linear dependence on $t_{\rm NS}$, and is around 0.1--1~s over most of the parameter space.  (The precision does not get much worse if we instead fit for $t_{\rm NS}$ and the exponent together.)  We expect this to be a powerful tool to distinguish between different PNS models.

$\bullet$
For early BH formation due to a failed supernova (Case 2), we assume that the event rate follows $1/t$ until it is sharply truncated at $t_{\rm BH}$.  The precision $\sigma = 1/R$, with $R$ being the event rate at $t_{\rm BH}$~\cite{Beacom:1998fj}; because we assume the event rate goes $1/t$, $\sigma$ then scales with $t_{\rm BH}$.  An extremely interesting scenario is if a BH forms between 0.1--1~s, as then Super-K could measure $t_{\rm BH}$ with $\simeq$1~ms precision.  This is comparable to the expected width of the BH formation transition.  Even if a BH forms at 1--10~s, Super-K might be able to probe the details of this transition if the core collapse is within 1~kpc (similarly for $\bar\nu_e$ in JUNO, or Hyper-K at 3~kpc).  No other detectors have this sensitivity.

$\bullet$
For late BH formation due to a metastable PNS (Case 3), we assume the event rate follows that of the NS case until it is truncated at $t_{\rm BH}$. Because the precision of the transition time is $\sigma = 1/R$, the shape of the curve is the inverse of our nominal luminosity. Even with this conservative assumption, the formation time can still be measured precisely.  For example, a BH forming at 40~s can be detected at 3$\sigma$ and $t_{\rm BH}$ can be measured to a precision of about 1~s.

Thus, if a NS is formed, it can be detected with high significance and the transparency time can be measured precisely in Super-K.  If a BH forms at early times, e.g., due to a failed supernova, Super-K or JUNO or especially Hyper-K may be able to probe the last instants of the transition to a BH.  Even if it forms at late times, the time of the transition can be measured precisely.  These measurements will powerfully test models.


\section{Conclusions}
\label{sec:conclusions}

Understanding core collapse is critical to astrophysics and physics, and detecting neutrinos is essential to progress.  The essential difficulty is that present neutrino detectors can only observe bursts from events in the Milky Way or its satellites.  Within the next few decades, there will likely be only one nearby core collapse~\cite{Reed:2005en, Diehl:2006cf, Keane:2008jj,  2011MNRAS.412.1473L, Adams:2013ana}, and it is essential that we maximize the scientific return from observing it.  This requires full coverage in flavors ($\nu_e$, $\bar{\nu}_e$, and $\nu_x$), high statistics for each (at least thousands of events), spectrum data (as broad as possible), and detections to very late times (at least tens of seconds).

We will soon have a promising complement of neutrino detectors to observe the next nearby core collapse.  The most important detectors are those at the 10-kton scale, as these have high yields of individually identifiable signal events, good event reconstruction, and low backgrounds.  For $\bar{\nu}_e$, the best detector will be Super-K (and JUNO, plus eventually the much larger Hyper-K).  For $\nu_e$, it will be DUNE, and for $\nu_x$, it will be JUNO.  It is important to assess how well these detectors could detect a core-collapse event relative to the criteria above, and especially to suggest ways they could improve their capabilities.  It is not clear if these detectors will run simultaneously for long enough or if there will be a group of successor detectors with complete flavor coverage.  We may have only one chance to get this right.

In this paper, we focus on neutrino detection and associated phenomenology from the start of the PNS-cooling phase to the very latest times.  We present the first comprehensive study (building on earlier work~\cite{Keil:95, Pons:99, Pons:01a, Roberts:12, Nakazato:2012qf, Mirizzi:2015eza, Camelio:17, Suwa:2019svl,  Nakazato:2020ogl}), providing a complete conceptual framework and calculating results for all flavors, emphasizing spectra, and employing detailed detection physics.  Without all flavors, one cannot test deleptonization, cooling versus accretion, and neutrino mixing.  Without spectra, one cannot break the degeneracy in measured rate between neutrino luminosity and average energy.  Without going to late times, one cannot measure the total releases in neutrino energy and lepton number.  Without realistic detector calculations, one cannot identify needed improvements.

To summarize our results for particular detectors, we find the following using the nominal PNS cooling model of Sec.~\ref{sec:theory}.  While adopting other models may change the quantitative details---in particular, including the effects of convection and nucleon-nucleon correlations would shorten emission times, though how much is uncertain---we expect the qualitative features to be similar.
\begin{itemize}
\item
{\bf Super-K:} Its prospects are near ideal, though it should seek to improve further.  Beyond 10 s, Super-K should be able to measure $\simeq$250 $\bar{\nu}_e$ events out to $\simeq$50 s, with excellent spectra before then.

\item
{\bf DUNE:} It has great potential, but significant new work is needed immediately, for which new resources are required.  Beyond 10 s, DUNE should be able to measure $\simeq$110 $\nu_e$ events out to $\simeq$40 s, with moderately good spectra before then.  However, detector backgrounds could become overwhelming at $\simeq$20~s, losing precious signal.  Increased efforts are needed on reducing backgrounds, energy resolution, particle identification, and other issues that affect signal sensitivity, especially near threshold.  Increased investments in light-detection systems, detector shielding, and data-acquisition electronics are needed.  The MeV neutrino-argon cross section has never been measured and is poorly known theoretically; dedicated measurements are feasible and needed.  All of these actions must be taken urgently.  Measuring a Milky Way core collapse is one of DUNE's three primary missions, and no other detector can measure $\nu_e$ well enough.

\item 
{\bf JUNO:} Its neutral-current channel potential is good, though it could be better, which is challenging.  (JUNO's $\bar\nu_e$ capabilities are comparable to those of Super-K.) Beyond 10 s, JUNO should be able to measure $\simeq$10 neutral-current events out to $\simeq$20~s, with fair spectra before then.  JUNO is making great progress on its neutral-current capabilities, with a key goal of lowering the threshold.

\end{itemize}
The experimental collaborations should better support phenomenological studies by providing more information about detector performance.  They should also publish dedicated late-time sensitivity studies themselves.

To summarize our results for the physics prospects, we expect that the detectors above can make critically important measurements.  For all flavors, they can measure the time evolution of the neutrino luminosities and average energies, which are sensitive probes of the underlying physics, including the equation of state of the PNS and neutrino interactions with the nuclear medium.  A key prediction is the eventual convergence of the emission from different flavors, which likely happens within several seconds after core collapse.  Complete measurements to late times are critical to measuring the total energy and lepton-number releases during NS formation.  Late-time measurements are also needed to distinguish between the PNS forming a NS or BH.  For NS formation, the PNS luminosity drops steeply as the neutrinospheres recede, which may happen in $\simeq$10--30 s.  For BH formation, the luminosity is suddenly truncated, which could happen over a wide range of times.  For the models we consider, Super-K can distinguish these outcomes out to several tens of seconds.  With improvements, DUNE could as well. For either the NS or BH case, the transition time can be measured with precision typically better than 1 s, which is also a sensitive probe of the underlying physics.

We are likely to have only one chance to make complete, precise measurements of a Milky Way core collapse.  If we are successful, the results will be an extraordinary Rosetta Stone for astronomy and physics, with wide-ranging impacts over decades.  If we are not, our understanding of core collapse and neutrinos may forever be subject to large uncertainties.



\section{Acknowledgments}

We thank Biswaranjan Behera, Francesco Capozzi, Erin Conley, Basudeb Dasgupta, Alexander Friedland, Christopher Grant, Alexander Himmel, Christopher Hirata, Charles Horowitz, Ivan Lepetic, Yufeng Li, Joseph McEwen, Alessandro Mirizzi, Payel Mukhopadhyay, Ken'ichiro Nakazato, Jos\'e Pons, Nirmal Raj, Sanjay Reddy, Ibrahim Safa, Kate Scholberg, Krzysztof Stanek, Hideyuki Suzuki, Irene Tamborra, Yun-Tse Tsai, Mark Vagins, Shun Zhou, and especially Bryce Littlejohn, Georg Raffelt, Michael Smy, Todd Thompson, Liangjian Wen, and Guanying Zhu for helpful discussions.

SWL and JFB were supported by NSF Grant No.\ PHY-1714479 awarded to JFB.  SWL was also supported by an Ohio State Presidential Fellowship, and later at SLAC by the Department of Energy under Contract No.\ DE-AC02-76SF00515. LFR was supported by the Department of Energy under Contract No.\ DE-SC0017955.



\begin{appendix}

\section{Supernova localization and distance}
\label{sec:distance}

Here we discuss methods to determine the direction and distance of a successful supernova with an optical display.  The distance is needed to turn neutrino fluxes into luminosities.  (For BH formation with no optical supernova, some estimate of the distance will be possible by assuming the total energy release, and it may be possible to detect the disappearance of the massive star~\cite{Kochanek:2008mp, Gerke:2014ooa, Adams:2016ffj}.) We first discuss finding the supernova, which is likely to be successful and will provide the distance measurement, and then comment on searches for the progenitor star.

The search for the optical supernova will likely be triggered by detection of the main neutrino signal, which precedes the supernova by $\simeq$0.1--1 days~\cite{Kistler:2012as, Nakamura:2016kkl} and which can determine its direction to a few degrees~\cite{Beacom:1998fj, Tomas:2003xn, Abe:2016waf, Brdar:2018zds, Hansen:2019giq, Linzer:2019swe}. For a nearby supernova, it may be possible to detect the pre-supernova neutrino emission, which would provide an alert~\cite{Odrzywolek:2003vn, Odrzywolek:2010zz,
Asakura:2015bga, Patton:2015sqt, Simpson:2019xwo, Raj:2019wpy}.

In Adams et al.~\cite{Adams:2013ana}, the prospects for optical detection of a Milky Way core-collapse supernova are detailed.  The distances containing 10\%, 50\%, and 90\% of supernovae in the galaxy are approximately 5, 10, and 15~kpc, a relatively narrow range.  (Even though the Large Magellanic Cloud, the largest of the Milky Way dwarf companions, hosted SN 1987A, this was an improbable event, as its expected supernova rate is $\simeq$ 10\% that of the Milky Way~\cite{Leahy:2016nfw, 2017ApJS..230....2B}.)  Since publication of Ref.~\cite{Adams:2013ana}, the distance to the Galactic Center has been revised from about 8.7 kpc to about 8 kpc~\cite{2016ApJ...832L..25H}, which we estimate reduces the median supernova distance to about 9 kpc.  Nevertheless, for easier comparison to other literature, we assume the standard distance of $D = 10$~kpc.

It is often assumed that astronomical surveys continually monitor the entire sky in optical light, including the Milky Way. Until recently, this has not been true to any useful depth, as most telescopes have narrow fields of view as a tradeoff for depth. That changed with the advent of the All-Sky Automated Survey for SuperNovae (ASAS-SN)~\cite{2014ApJ...788...48S, 2017PASP..129j4502K}, which is presently operating 20 robotic telescopes at four sites worldwide, allowing nearly weather-proof monitoring of the full available sky—including the Milky Way plane—every 24 hours (except for the approximately 25\% of the sky toward the Sun). Each telescope has a field of view of $4.5 \times 4.5$ square degrees and an angular resolution of 16”. In routine observations, ASAS-SN reaches a depth $g \sim 18$ in $3 \times 90$~s dithered exposures. ASAS-SN discovers or recovers $\simeq$500 bright extragalactic supernova per year~\cite{Holoien:2017sbn, Vallely:2019kra}, also discovering a plethora of other transients (e.g., Refs.~\cite{2019ApJ...876..115S, Holoien:2019zry}) as well as many thousands of variable stars (e.g., Refs.~\cite{2018MNRAS.477.3145J, 2020MNRAS.491...13J}).

For a Milky Way supernova, ASAS-SN has a high detection probability.  Reference~\cite{Adams:2013ana} shows the fraction of supernovae that can be observed at a given depth.  Results are not shown for $g$-band, but they are shown for $V$-band, which was formerly used by ASAS-SN.  In $V$-band, routine observations ($V \sim 17$) would capture $\simeq$90\% of core-collapse supernovae, though the uncertainties on the dust-extinction maps may not be negligible.  Deeper observations using a crude direction from neutrino pointing could go $\sim 2$ magnitudes deeper, which would instead capture $\simeq$95\%.  While the $g$-band now used by ASAS-SN is in fact bluer than $V$-band, with nominally worse dust extinction, the photometric depth achieved is significantly higher, so we expect the current ASAS-SN configuration to have even better recovery fractions.

In addition to ASAS-SN, a wide variety of telescopes across the full electromagnetic spectrum would certainly be quickly marshaled to search for the supernova.  Many bands are insensitive to extinction by dust, and some can be used even in directions near the Sun (e.g., MeV gamma rays~\cite{Wang:2019zsk}). In many bands, the supernova will remain detectable for months or even years~\cite{2014ApJ...792...10S, 2016ARA&A..54...19M}, allowing a wide range of techniques for estimating distances to be used~\cite{Arnett:1990au}.  How well the distance can be determined depends on details that cannot be known yet.  For SN 1987A, the distance was ultimately determined to $\simeq$2\%~\cite{1999IAUS..190..549P}.  For a future Milky Way supernova, the closer distance and better instrumentation will provide advantages, though there will be challenges due to the supernova not being in an isolated host galaxy, as SN 1987A was in the Large Magellanic Cloud.

Once the supernova has been detected, there will also be extensive searches for the progenitor star in pre-explosion imaging, as has been successful for a good number of supernovae in nearby galaxies~\cite{Smartt:2009, Smartt:2015sfa}.  The prospects for detecting the progenitor star and for isolating it from nearby stars are discussed in Ref.~\cite{Adams:2013ana}.  If successful, these measurements of the progenitor magnitude and perhaps temperature may further constrain the distance to the supernova.

\end{appendix}


%


\end{document}